\documentclass[12pt,preprint]{aastex}

\def\lea{\mathrel{<\kern-1.0em\lower0.9ex\hbox{$\sim$}}}
\def\gea{\mathrel{>\kern-1.0em\lower0.9ex\hbox{$\sim$}}}

\slugcomment{Accepted for publication in The Astronomical Journal}

\shorttitle{M33 Star Cluster Catalog}
\shortauthors{Sarajedini & Mancone}

\begin{document}


\title{A Catalog of Star Cluster Candidates in M33}

\author{Ata Sarajedini and Conor L. Mancone}
\affil{Department of Astronomy, University of Florida, 211 Bryant Space
Science Center, Gainesville, FL 32611-2055}



\begin{abstract}
We present a new catalog of star cluster candidates in the nearby spiral
galaxy M33. It is based on eight existing catalogs wherein we
have cross-referenced identifications and endeavored to resolve inconsistencies
between them. Our catalog contains 451 candidates of which 255 
are confirmed clusters based on HST and high resolution
ground-based imaging. The catalog contains precise cluster positions (RA and Dec),
magnitudes and colors in the UBVRIJHK$_S$ filters, metallicities, radial velocities, 
masses and ages, where available, and galactocentric distances for each
cluster.  The color distribution of the M33 clusters appears to be similar to those in the
Large Magellanic Cloud with major peaks at $(B-V)_{0}$$\sim$0.15, and 
$(B-V)_{0}$$\sim$0.65. The intrinsic colors are correlated with cluster ages, which 
range from $10^{7.5}$ to $10^{10.3}$ years. The age distribution of the star clusters 
supports the notion of rapid cluster disruption with a slope of
$\alpha$=--1.09$\pm$0.07 in the $dN_{cluster}/d$$\tau$~$\propto$~$\tau$$^{\alpha}$ 
relation. In addition, comparison to theoretical
single stellar population models suggests the presence of an age-metallicity
relation among these clusters with younger clusters being more metal-rich.
Analysis of the radial distribution of the clusters yields some evidence that younger clusters 
(age $\lea$ 1 Gyr) may be more concentrated toward the center of M33
than older ones. A similar comparison 
with the radial profile of the M33 field stars shows the clusters to be
more centrally concentrated at the greater than 99.9\% confidence level. Possible
reasons for this are presented and discussed; however, the overwhelming conclusion
seems to be that a more complete and thorough cluster search is needed covering at
least 4 square degrees centered on M33.
\end{abstract}



\section{Introduction}


The identification of star clusters in M33 can be traced back to the pioneering work
of Hiltner (1960, hereafter Hilt), who used photographic plates taken with the Mt. Wilson
100-inch telescope to photometer 23 cluster candidates in the UBV passbands. 
He concluded that the clusters in M33 are generally bluer and fainter than those in M31. 
The next major catalog was published by Melnick \& D'Odorico (1978,
hereafter MD) adding 33
more objects to the census of star cluster candidates. Their assertion that
M33 seemed to contain too many globular clusters for its luminosity led them
to conclude that some of the cluster candidates are associated with the
disk of M33. The most comprehensive catalog of non-stellar objects in M33
was compiled by Christian \& Schommer (1982, hereafter CS) using
a single photographic plate taken at the Ritchey-Chr\'{e}tien focus of
the 4m telescope at Kitt Peak National Observatory. Additional supporting
observational material was used to arrive at the final list of 250 objects in
the catalog. Subsequent papers analyzed the photometric, spectroscopic, and
kinematical properties of these clusters (Christian \& Schommer 1983; 1988;
Schommer et al. 1991). The most recent attempt to compile a catalog of
M33 clusters using ground-based facilities  is that of Mochejska et al. (1998, hereafter
MKKSS),
wherein 35 new cluster candidates were cataloged and 16 previously known
ones were confirmed. In addition to the cluster census, MKKSS
also presented an analysis of the M33 cluster color-magnitude diagram, color-color
diagram, and luminosity function as compared with the Milky Way.

The era of using space-based telescopes such as the Hubble Space
Telescope (HST) to identify M33 clusters began with the work 
of Chandar, Bianchi, \& Ford (1999, hereafter CBF99). They used images taken 
with the Wide
Field Planetary Camera 2 (WFPC2) aboard HST to identify 60 star clusters, 11 of which
were previously cataloged as nonstellar objects from ground-based
surveys. This was augmented by an additional set of 102 star clusters, 82 of
which were previously unknown,
presented by Chandar, Bianchi, \& Ford (2001, hereafter CBF01) again using the 
WFPC2
instrument. Both studies present positions for the clusters as well as integrated 
photometry in a variety of filters. Most recently, demonstrating the power of
the Advanced Camera for Surveys (ACS) Wide Field Channel
on HST for studies such as this, Bedin
et al. (2005, hereafter BEA) detect 33 star clusters and 51 candidates in one M33 field.
Sarajedini et al. (2007, hereafter SBGHS) have also used the
resolving power of ACS on HST to identify 24 star clusters of which 12 are previously
uncataloged. They demonstrate that the construction of cluster color-magnitude
diagrams provides powerful inputs into the interpretation of the
integrated-light properties. 

Alongside these catalogs, a number of papers led by Jun Ma have
been published on the properties of M33 clusters in the above-mentioned
catalogs (Ma et al. 2001, 2002a, 2002b, 2002c,
2004a, 2004b) including integrated magnitudes, colors, ages, masses, 
and metallicities. Using the Beijing-Arizona-Taiwan-Connecticut (BATC) 
filter system, the series of papers by
Ma et al. construct spectral energy distributions (SED) of known M33 cluster candidates
and use the shape of the SEDs to estimate cluster properties. 

While the proliferation of M33 cluster catalogs and the supporting work
by Ma et al. have been quite valuable, it is clear that a single master catalog 
incorporating the entries in all of the individual catalogs including all 
known properties of each cluster would be an important step forward.
Constructing such a catalog of M33 star
clusters has a number of advantages. First, it provides a standard
positional reference frame and photometric zeropoint  for future catalogs. 
Second, having a catalog that contains ALL previous catalogs plus cluster
properties is important in helping us to better understand the M33 cluster
system and M33 itself. 

Throughout this paper, we make a distinction between
the full version of our catalog available via the world wide web
\footnote{http://www.astro.ufl.edu/$\sim$ata/cgi-bin/m33\_cluster\_catalog/} 
(FC for full catalog) and the cluster catalog of adopted values included in 
the present work (AC for adopted catalog). The former contains
the properties of each cluster as quoted in all of the referenced works. The
latter, which is analyzed in this manuscript, contains only our adopted values for
such parameters as the cluster photometry, age, and mass.
The next section is a brief overview of the catalogs
that we have used. Section 3 describes in detail the construction of this 
new catalog and Section 4 includes an analysis of the cluster properties.
Lastly, our conclusions are presented in Section 5. 

\section{Existing Catalogs}

In Sec. 1, we noted the 8 cluster catalogs (Hilt, MD, CS, MKKSS, CBF99, CBF01, BEA,
and SBGHS) and 6 papers containing cluster properties (Ma01, Ma02a, Ma02b, Ma02c,
Ma04a, Ma04b) that we plan to integrate into our new catalog of M33 cluster data. Table 1
lists the bibliographic citation of each source along with the abbreviation we will
use in the present paper. Table 1 also lists the information contained in each of these sources. 
Our primary sources for cluster identifications are Hilt, MD, CS, MKKSS, CBF99, CBF01, 
BEA, and SBGHS. Some of these papers also provide photometric measurements.
Cluster properties such as ages, masses, and metallicities are taken from the
Ma et al. series of papers. In particular, Ma01 and Ma02b present properties
for CBF99 and CBF01 clusters, Ma02a and Ma04b provide additional data for the MD 
clusters, Ma02c presents ages for clusters identified by MKKSS, and Ma04a gives
metallicities for the old star clusters in M33.

\section{New Catalog}

\subsection{Cluster Positions}

All of the input catalogs provide right ascensions and declinations for the clusters
except for Hilt and MD, which only provide finder charts. 
The positions of the clusters were transformed to
the J2000 epoch and refined using the Local Group Survey (LGS, Massey et al. 2006)
images of M33 available from their ftp site
\footnote{ftp://ftp.lowell.edu/pub/massey/lgsurvey/datarelease/}. These are NOAO 
MOSAIC frames of 3 overlapping fields in M33 that have been registered and stacked
to yield combined UBVRI images. The IRAF task 
\verb/imexamine/ was used to determine the cluster positions on the V frames
and \verb/wcstran/ was used to reference them to the World Coordinate System of each image.
The positions
are relative to the USNO-A2.0 catalog and have a rms error of $\sim$0.25 arcsec. These
are the positions that are used in the FC and AC versions of the catalog.
We note that three clusters (SM 442, SM 450, and SM 451)
fell outside of the region covered by the LGS images. In these cases,
we measure the cluster positions on images taken from the Digitized Sky Survey.
In most cases, the position listed for a given cluster in the original catalog was of sufficient
accuracy to make the cluster location easily discernable.  
In crowded regions or for faint clusters, the cluster's location on the LGS image was confirmed by 
referring to the images used in the original paper - typically HST/WFPC2 frames as
in the work of CBF99 and CBF01.  In the case of the Hilt and MD catalogs,
the finder charts were used exclusively to locate the clusters.

The Christian \& Schommer (1982) cluster positions and identifications deserve further
discussion. Their right ascensions and declinations are only accurate to about 20 arc seconds, 
so the CS charts were used in most cases to confirm the identity of the clusters. Furthermore,
 in their original catalog 
CS listed 18 miscellaneous objects but did not include positions for them.  Three of these objects 
(M9, M11, M12) were labeled on their finding chart and have been included in the present catalog.  
Eight of these objects (M1, M2, M4, M5, M6, M8, M10, M15) were listed with cross-identifications to 
MD.  These cross-identifications were assumed to be correct and the CS identifications have been
 added to our catalog.  
The remaining objects (M3, M7, M13, M14, M16, M17, M18) are currently unidentified and were 
not included in our catalog.  

Figure 1 shows the offsets in right ascension and declination between our positions
derived from the LGS images and the positions listed in each individual catalog. The 
sense of the difference is given as (This work -- Others). It is clear from Fig. 1 that the
root-mean-square deviations of the offsets are all quite small - less than $\sim$1 arcsec, with
the exception of CS, which is closer to $\sim$10 arcsec. This is consistent with the astrometric
precision claimed by CS for their positions.

\subsection{Cross Identifications}

Using the measured positions from the Local Group Survey images, we cross-identified the 
various catalogs with each other.  Any two clusters located within 
0.25 arc seconds of each other were assumed to be the same cluster.  
When two or more matching clusters were found, they were considered one entry in
the catalog with one position but the photometry and other cluster properties from 
all available sources are kept and stored as part of the FC entry.  The original papers 
listed a total of 608 clusters.  When these are combined into one catalog, 451 unique
objects emerge.  Of these 451 cluster candidates, 105 of them appeared in more than 
one catalog source, not including 
the Ma et al. papers, which give cluster properties rather than newly identified clusters.  
In addition, there are 4 clusters in the CBF compilation that appear to be duplicates
based on our position-matching algorithm: CBF99-22 = CBF01-91, CBF99-15 = CBF99-45,
CBF99-56 = CBF01-156, CBF99-60 = CBF01-94. These have also been noted in the FC 
version of our database. 

Of the 451 objects in our final catalog only 203 of these have been imaged with HST and 
can be confidently declared clusters.  These represent a combination of WFPC2 images
used in CBF99 and CBF01, ACS observations used by BEA, and Near-Infrared Camera
Multi-Object Spectrograph (NICMOS) and Space Telescope Imaging Spectrograph
(STIS) images we extracted from the HST archive to classify candidates in our
catalog. 
The remaining 248 objects are likely a 
combination of clusters, galaxies, HII regions, and perhaps other stellar aggregates.  In order to 
minimize this possibly significant source of contamination in our catalog, we made use
of archival M33 images taken with the MegaPrime/MegaCam instrument
on the Canada-France-Hawaii Telescope (CFHT)
under excellent seeing conditions ($\sim$0.5").  Every object that wasn't observed with 
HST was visually inspected on the CFHT images.  Objects were divided into 
5 categories: clusters, galaxies, stars, unknown, and 
objects that fell in a gap between the CCD chips that constitute the MegaPrime imager.  
Of the 248 objects without HST imaging, only 52 were classified as clusters.  Combined, the 
203 HST clusters and 52 ground based clusters form the high confidence set of 255 clusters used 
in the discussion section below.

\subsection{Photometric Standardization}

The photometry from the various original catalogs are all on different zeropoints. As such,
we have adopted one of the catalogs as our photometric standard and offset all
of the other catalogs to this standard. Because it contains an extensive set of CCD
photometry in multiple filters, we have decided to use the CBF photometric scale
as our standard. The photometry from each catalog was compared with that of CBF
and an offset was calculated using a 2-$\sigma$ rejection algorithm. Table 2 gives
the values of these offsets, in the sense (Catalog--CBF), along with the standard 
deviations and standard errors of the means.
Note that only the CS catalog contained R magnitudes so these were not
transformed in any way. In addition, the U magnitudes are almost exclusively from CBF 
because although Hiltner provides U mags, there
is only one cluster in common between them. There are no clusters in common between
CBF and Hiltner which have B mags, and all but two of the clusters were measured by
MD. As a result, we have ignored the U and B photometry from Hiltner.

Figures 2 through 4 illustrate the magnitude differences in B, V,
and I as a function of V and B--V or V--I between each input catalog with photometry and that
of CBF. Inspection of these plots reveals no apparent systematic trends in the
magnitude differences with magnitude or color. In addition, the scatter about the
mean is generally similar for all of the catalogs except for the Ma et al. photometry,
which displays the greatest dispersion about the mean as shown in Table 2.
This is probably due to the fact that the original photometry presented in the
Ma et al. series of papers was obtained in the proprietary BATC filters
and transformed to the BVI system using standard stars from Landolt (1983;
1992) as described by Ma02a and Ma02b.
However, it is important to note that the standard error of the means for the Ma et al.
photometry is not significantly higher than for the other catalogs.

Our final adopted magnitudes are the average of all corrected measurements excluding
the Ma et al. values.  When other photometry was available, the Ma et al. values were excluded 
from our final results because of their apparently larger errors.  In 6 cases 
(MD 2, MD 18, MD 32, MD 33, MD 41, MD 44) only Ma et al. provide V magnitudes, so we 
adopted their corrected photometry for these clusters.
We have supplemented these optical magnitudes with near-infrared $JHK_S$ photometry
from the point source catalog of the Two Micron All Sky Survey (2MASS)
\footnote{See http://irsa.ipac.caltech.edu/}. Eighty-five of the cluster candidates in our
catalog possess 2MASS photometry. 

The adopted catalog of cluster properties is given in Table 3. For each cluster, we
list the identification number, RA and Dec in the J2000 epoch, V, B--V, V--I on the
CBF photometric system, the logarithms of the age in years and mass in solar masses,
along with a classification - cluster, stellar, unknown, galaxy - and alternate bibliographic
sources where the cluster appears. The properties of the confirmed clusters in
this sample are analyzed and discussed in the next section.

\section{Results and Discussion}

Now that we have assembled our cluster compilation, we are in a position to analyze the properties
of the clusters themselves. The two panels of Fig. 5 show the color-magnitude 
diagrams (CMDs) for the 255 high-confidence star clusters in M33 and 501 star
clusters in the  Large Magellanic Cloud (LMC) from Bica et al. (1999).  
Note that we have not included the entries in the Bica et al.
(1999) catalog identified as `associations.' All colors have been dereddened with a uniform 
value of E(B--V)=0.1, as typical of the published values for the line-of-sight reddenings to
M33 and the LMC.
We adopt a distance modulus of $(m-M)_0 = 24.69$
(Galleti et al. 2004) for M33 and $(m-M)_0 = 18.40$ (Grocholski et al. 2007) for the LMC.

The most striking difference between the M33 and LMC cluster CMDs is that 
the latter population extends to as faint as $M_V$$\sim$--4.0 while the M33 clusters
terminate at a point 1.5 mag brighter. This may suggest that our M33 cluster catalog represents a
photometrically incomplete sample. However, this possibility
can only be addressed with a deeper and more extensive 
homogeneous imaging survey of M33.
The lower panel of Fig. 5 illustrates the color distribution of the
M33 and LMC clusters scaled to unit area. We see that both galaxies exhibit distinct
cluster populations with $(B-V)_{0}^{peak}$$\sim$0.15, and $(B-V)_{0}^{peak}$$\sim$0.65. 

The colors of the clusters appear to be strongly  correlated with their ages as illustrated in
Fig. 6. We begin by noting that Fig. 6a plots the absolute magnitudes of the M33 clusters
as a function of their ages all of which come from the Ma et al. series of papers. The
solid lines represent single stellar population models with Z = 0.004 and 
masses of 10\textsuperscript{2}, 10\textsuperscript{3}, 10\textsuperscript{4}, 
10\textsuperscript{5}, and 10\textsuperscript{6}$M_{\odot}$ from Girardi et al. (2002) 
adopting a mass-to-light ratio of unity. We can use these model loci to calculate a mass for
each cluster and compare that with their ages. This is shown in Fig. 6b.
We see that there is a tight correlation between cluster mass and age with older clusters 
having preferentially higher masses. This is highly reminiscent of what is seen 
among the star clusters in the Large and Small Magellanic Clouds (Hunter et al. 2003).
We note that the lower mass envelope of this relation is undoubtedly due to the
fading of  clusters over time. In fact, the solid line
represents the fading line predicted by the Bruzual \& Charlot (2003) models for Z=0.008
shifted to match the lower envelope of points. The upper envelope of the points in Fig. 6b
is likely a result of the so-called `size-of-sample' effect as described in Hunter et al. (2003) and
Whitmore, Chandar, \& Fall (2007). Figure 6c illustrates the 
relation between dereddened color and cluster age. Once again, there is a good correlation
between cluster color and age with older clusters being redder.
The lines represent single stellar
population models from Girardi et al. (2002) for a low metallicity (Z=0.0004, dashed) 
and the solar value (Z=0.019, solid). We see that at old ages, the data points are 
more consistent with the metal-poor model while at younger ages, they are closer
to the solar abundance model. This suggests the presence of a significant 
age-metallicity relation among the M33 clusters. 

We plot the age distribution of star clusters in M33 in Figure 7.  The number of 
clusters appears to decline with age with no obvious breaks or abrupt changes.  
Following Fall et al. (2005) and Chandar et al. (2006), we fit a power law of the form 
$dN_{cluster}/d$$\tau$~$\propto$~$\tau$$^{\alpha}$, and find $\alpha$=--1.09$\pm$0.07.  
Although the completeness of the M33 cluster sample is likely quite complicated, Figure 6b 
suggests that our sample is approximately luminosity limited.  The results are similar 
to the slope of $\sim-1.1$ found by Rafelski \& Zaritsky (2005) for clusters in the SMC.


Next, we explore the radial variation of the cluster ages. The top and bottom panels of 
Fig. 8 display the dereddened color and age of each cluster, respectively, 
as a function of deprojected galactocentric radius. We have adopted 
$\alpha_{J2000}$=23$^h$ 27$^m$ 45$^s$, 
$\delta_{J2000}$ = 30$^o$ 39' 36" for the center of M33, and the
deprojection has been calculated 
using the position angle (23$^o$) and inclination (56$^o$) provided by Regan \& Vogel (1994). 
Both panels of Fig. 8 suggest that bluer (younger) clusters are more centrally
concentrated as compared with redder (older) clusters. This difference is better 
investigated using the cumulative radial distributions of the two populations as
illustrated in Fig. 9 and an application of the Kolmogorov-Smirnov (K-S) test.
The solid lines in Fig. 9 show the cumulative radial positions of
the 255 confirmed clusters in our catalog with the black line representing all clusters,
the blue line showing just the blue clusters [$(B-V)_0$$<$0.5], and the red line for the 
red clusters [$(B-V)_0$$>$0.5]. Division of the clusters at a color of $(B-V)_0$$=$0.5
represents an age of $\sim$1 Gyr (see Fig. 6). There is no reason to believe that the completeness
of our catalog varies with cluster color, so we proceed to apply the K-S test to
the solid red and blue distributions in Fig. 9. We see that the blue clusters are more 
centrally concentrated than the red clusters
at the 88\% significance level. Though not significant at the $>$95\% level, this
result is suggestive and worthy of rexamination once a larger sample of M33 clusters
becomes available.

We now seek to examine the radial density distribution of our cluster sample. The
filled circles in Fig. 10 show the cluster density profile with the upper panel plotting deprojected
radius and the lower panel showing projected radius. Radii in arcminutes and kiloparsecs
are given using our adopted distance modulus of $(m-M)_0=24.69$. Inside $\sim$10
arcmin, the cluster profile exhibits a flat density distribution with occasional dips that
probably suggest some level of incompleteness. Outside of $\sim$10 arcmin, the behavior
is essentially a power law with the most distant clusters located at a distance of 
$\sim$29 arcmin or $\sim$7.2 kpc from the center of M33 in projected distance. 
This decrease could 
represent the genuine `edge' of the cluster distribution or it could be a result of
radial incompleteness in all previous M33 cluster censuses. For the discussion
below, we proceed under the assumption that this decrease in cluster density
at large radii has not been adversely affected by the shortcomings of previous 
cluster catalogs.

It is important to place the cluster density distribution within the context of the
field stars in M33. To expedite this, we make use of the stellar catalog provided by
the ``M33 CFHT Variability Survey" of Hartman et al. (2006). This catalog contains
multi-color photometry for 4.7 million point sources in a 1 square degree field centered
on M33 from the MegaPrime/MegaCam instrument on the CFHT. 
The color-magnitude diagrams published by Hartman et al. (2006) extend to a
magnitude limit of i'$\sim$24.5 with photometry in the Sloan g', r', and i' filters.
The solid lines in Fig. 10 represent the radial density distribution of the field stars
from the Hartman et al. (2006) survey compared to the high-confidence M33 star clusters in the
present catalog. The stellar density distribution has been scaled to match the cluster
density in the inner-most radial bin. 

Figure 10 shows that the stars in M33 exhibit a much larger radial extent
than the clusters. At a given cluster density, the stars extend between 2 and 5 kpc
beyond the clusters in deprojected distance. This impression is borne out by the
application of the K-S test to the two distributions (Fig. 9); there is a greater than 99.9\% 
chance that the stars and clusters are drawn from different parent populations. 
However, we need to be cognizant of
the possibility that the cluster and stellar samples may have different completeness
properties. For example, both the stellar and cluster distributions show signs of
incompleteness toward the center of M33. The cluster profile flattens out and shows 
uncharacteristic dips inside of 10 arcmin from the galaxy's center while the stellar density 
profile actually decreases and exhibits a negative radial slope inside 10 arcmin. In order
to minimize the influence of potential incompleteness in these samples, we can
limit the comparisons to objects outside of 10 arcmin from the center of M33. At these
radii, the cluster and stellar distributions have a better chance of possessing similar 
completeness properties. However, even when we limit our comparison
to these subsamples, there is still a greater than 99.9\% chance that the stars
and clusters are drawn from different  populations. 

If this difference between the stellar and cluster radial profiles is a genuine astrophysical
phenomenon and not the result of observational biases in the samples,
then there are a number of possible explanations for it. First, there is the process of
orbital diffusion which, over time, increases the mean galactocentric distance of a 
population as a result of gravitational interactions with more massive objects such as
giant molecular clouds (Wielen 1977; Wielen, Fuchs, \& Dettbarn 1996). 
In this scenario, individual stars, being much less massive than
star clusters, are more susceptible to orbital diffusion so that they are more
likely to be located at larger galactocentric distances as compared with clusters.
In fact, the work of Carraro \& Chiosi (1994) suggests that even low mass
stellar systems such as Milky Way open clusters are minimally affected by
orbital diffusion. To test the effect of orbital diffusion, we have divided up the 
stellar sample into two age groups - 
those with colors representative of young main sequence stars (age$\lea$300 Myr)
and those on the first ascent red giant branch (age$\gea$3 Gyr). Figure 9 shows
a comparison of the cumulative radial distributions of these groups. We find a
K-S probability of greater than 99.9\% that the blue (younger) stars are more centrally
concentrated than the red (older) stars. This could be the result of orbital diffusion,
which will affect the older stars to a greater degree than the younger stars, but
this difference could simply be
due to the fact that that the higher gas densities at smaller radii have resulted in more 
recent star formation. As a result, whether the process of orbital diffusion is largely
or partially responsible for the greater radial extent of the stars as compared to
the clusters in still an open question.

Another possible explanation for the difference between the cluster and stellar
profiles in Fig. 10 is that at the lower gas densities of the outer regions of M33, stars
or small groups of stars are more likely to form than larger more massive clusters
(Tasker \& Bryan 2006, 2007).
In this case, we should be able to detect a radial gradient in the mean masses
of the clusters with lower mass clusters being present at larger galactocentric radii.
Such a diagram has been constructed using our cluster catalog, but no significant
trend is apparent. In any case, if the result that the field stars in M33 exhibit a 
significantly greater radial extent than the clusters holds up to further scrutiny, 
it could have important consequences for our understanding of M33's
star formation and dynamical history.

\section{Summary}

We have combined eight published catalogs of star clusters in M33 into one 
coherent database with accurate right ascensions and declinations measured 
from the Local Group Survey images of Massey et al. (2006). This catalog contains
451 cluster candidates of which 255 are confirmed based on HST and high resolution
ground-based imaging. The catalog also contains magnitudes and colors in the 
UBVRIJHK$_S$ filters on a consistent photometric system. In addition, we have included such
information as cluster metallicitiies, radial velocities, masses and ages as well as 
galactocentric distances in the catalog. 

The color-magnitude diagram of the M33 star clusters shows integrated magnitudes
in the range --9$\lea$$M_V$$\lea$--4.5 and colors of --0.5$\lea$$(B-V)_0$$\lea$1.0.
 The color distribution of the M33 clusters appears to be similar to those in the
 LMC with major peaks at $(B-V)_{0}$$\sim$0.15, and $(B-V)_{0}$$\sim$0.65.
The intrinsic colors of the M33 clusters are correlated with their ages, which 
range from $10^{7.5}$ to $10^{10.3}$ years. In addition, comparison to theoretical
single stellar population models suggests the presence of an age-metallicity
relation among these clusters with younger clusters being more metal-rich.

Analysis of the radial distribution of the clusters suggests that younger clusters 
(age $\lea$ 1 Gyr) may be more centrally concentrated than older ones, though the
statistical significance of this result is only at the 88\% level. A similar comparison
with the radial profile of the M33 field stars however shows the clusters to be
more centrally concentrated at the greater than 99.9\% confidence level. Possible
reasons for this are presented and discussed; however, the overwhelming conclusion
seems to be that a more complete and thorough cluster search is needed covering at
least 4 square degrees centered on M33.

\acknowledgments
The authors wish to thank Mike Barker for assisting with some of the data
gathering for this project as well as many stimulating discussions. 
Michael Fall, Jonathan Tan and Elizabeth Tasker provided a number of intriguing
ideas in the process of interpreting these data. In addition, we are grateful to
Rupali Chandar for a number suggestions that greatly improved the catalog and
this manuscript. This research was funded by NSF CAREER grant AST-0094048 to A.S.

\clearpage


\begin{center}
\tabletypesize{\tiny}
\begin{deluxetable}{lcccccccccccccc}
\rotate
\tablewidth{0pt}
\tablecaption{Bibliographic Sources and their Contents}
\tablehead{
\colhead{Source} & \colhead{Abbreviation} & \colhead{Position}    &
\colhead{V}     & \colhead{B}           &
\colhead{I}     & \colhead{F775W}       &
\colhead{B--V}  & \colhead{U--B}                &
\colhead{U--V}  & \colhead{V--I}                &
\colhead{V--R}  & \colhead{Age}         &
\colhead{Mass}  & \colhead{[Fe/H]}}
\startdata
Hiltner (1960)& Hilt & \nodata &x&  \nodata &  \nodata &  \nodata &x&x&  \nodata &  \nodata & \nodata  &  \nodata & \nodata  &  \nodata \\
Melnick and D'Odorico (1978)& MD &  \nodata &  \nodata &x&  \nodata &  \nodata &  \nodata & \nodata  &  \nodata & \nodata  &  \nodata &  \nodata &  \nodata &  \nodata \\
Christian \& Schommer (1982)& CS& x&x& \nodata  &  \nodata & \nodata  &x&  \nodata &  \nodata & &x& \nodata  & \nodata  &  \nodata \\
Christian \& Schommer (1988)& \nodata  & \nodata  &x&  \nodata &  \nodata &  \nodata &x& \nodata  & &x& \nodata  &  \nodata &  \nodata &  \nodata \\
Mochejska et al. (1998)& MKKSS &x&x&x&x& \nodata  &x& \nodata  & \nodata  &x& \nodata  & \nodata  & \nodata  & \nodata  \\
Chandar, Bianchi, \& Ford (1999)& CBF99 &x&x& \nodata  &  \nodata & \nodata  &x&x&x&  \nodata &  \nodata &  \nodata & \nodata  & \nodata  \\
Chandar, Bianchi, \& Ford (2001)& CBF01 &x&x& \nodata  & \nodata  & \nodata  &  \nodata &  \nodata & &x& \nodata  &  \nodata & \nodata  &  \nodata \\
Bedin et al. (2005)& BEA &x& \nodata  & \nodata  & \nodata  &x&  \nodata &  \nodata & \nodata  &  \nodata &  \nodata &  \nodata & \nodata  & \nodata  \\
Sarajedini et al. (2006)& SBGHS &x&x&  \nodata & \nodata  & \nodata  & \nodata  & \nodata  & \nodata  &x& \nodata  & \nodata  & \nodata  & \nodata  \\
Ma et. al (2001)& Ma01 &  \nodata &x& \nodata  & \nodata  & \nodata  &x& \nodata  & \nodata  & \nodata  & \nodata  &x& \nodata  &x\\
Ma et. al (2002a)& Ma02a &x&x& \nodata  &  \nodata &  \nodata &x& \nodata  & \nodata  & \nodata  & \nodata  & \nodata  & \nodata  & \nodata  \\
Ma et. al (2002b)& Ma02b & \nodata  &x& \nodata  & \nodata  & \nodata  & \nodata  & \nodata  & \nodata  & \nodata  & \nodata  &x& \nodata  &x\\
Ma et. al (2002c)& Ma02c & \nodata  &x& \nodata  & \nodata  & \nodata  &x& \nodata  & \nodata  & \nodata  & \nodata  &x& \nodata  &x\\
Ma et. al (2004a)& Ma04a & \nodata  & \nodata  & \nodata  & \nodata  & \nodata  & \nodata & \nodata  & \nodata  & \nodata & \nodata  &x& \nodata  &x\\
Ma et. al (2004b)& Ma04b & \nodata  & \nodata  & \nodata  & \nodata  & \nodata  & \nodata  & \nodata  & \nodata  & \nodata  & \nodata  &x&x& \nodata  \\
\enddata
\end{deluxetable}
\end{center}

\clearpage

\begin{deluxetable}{lcccccccccc}
\tablecaption{Photometric Offsets Relative to CBF}
\tablewidth{0pt}
\tablehead{
\colhead{Source}	& 
\colhead{$\Delta$B}	& $\sigma$  &   $\sigma$/$\sqrt{N}$ &
\colhead{$\Delta$V}	& $\sigma$  &   $\sigma$/$\sqrt{N}$ &
\colhead{$\Delta$I}  & $\sigma$  &   $\sigma$/$\sqrt{N}$ }
\startdata
SBGHS& \nodata & \nodata & \nodata & 0.031 & 0.233 & 0.070 & --0.020 & 0.294 & 0.088\\
MKKSS&--0.116   & 0.096   & 0.036     & --0.093 & 0.202 & 0.050 & 0.022 & 0.174 & 0.071 \\
CS& --0.177 & 0.165 & 0.096 & --0.015 & 0.208 & 0.043 &--0.030 & 0.632 & 0.316 \\
Hilt& \nodata & \nodata & \nodata & --0.073 & 0.078 & 0.029 & \nodata & \nodata & \nodata \\
MD& --0.023 & 0.083 & 0.041 & \nodata & \nodata & \nodata & \nodata & \nodata & \nodata \\
Ma& --0.060 & 0.249 & 0.038 & 0.043 & 0.306 & 0.027 & \nodata & \nodata & \nodata \\
\enddata
\end{deluxetable}

\clearpage
\begin{deluxetable}{cccccccccc}
\tabletypesize{\tiny}
\rotate
\tablewidth{0pt}
\tablecaption{M33 Adopted Cluster Catalog}
\tablehead{
\colhead{ID}  & \colhead{RA (J2000)} &
\colhead{Dec} & \colhead{V}  &
\colhead{(B--V)}   & \colhead{(V--I)}  &
\colhead{Log Age \tablenotemark{a}}  & \colhead{Log Mass \tablenotemark{b}} &
\colhead{Classification}   & \colhead{Alternate Source(s)}
}
\startdata
1&01:32:31.97&30:37:37.5&\nodata&\nodata&\nodata&\nodata&\nodata&Unknown&CS U92\\
2&01:32:33.36&30:26:20.9&\nodata&\nodata&\nodata&\nodata&\nodata&Stellar&CS H39\\
3&01:32:34.40&30:37:42.6&19.86&\nodata&1.13&\nodata&\nodata&Cluster&CBF 143\\
4&01:32:35.60&30:41:28.0&\nodata&\nodata&\nodata&\nodata&\nodata&Unknown&CS H22\\
5&01:32:38.87&30:47:07.1&\nodata&\nodata&\nodata&\nodata&\nodata&Galaxy&CS U33\\
6&01:32:38.97&30:39:17.9&19.92&\nodata&\nodata&10.28&\nodata&Cluster&CBF 162; Ma 2002b\\
7&01:32:39.13&30:40:42.0&\nodata&\nodata&\nodata&\nodata&\nodata&Unknown&CS U81\\
8&01:32:41.27&30:27:51.9&\nodata&\nodata&\nodata&\nodata&\nodata&Unknown&CS U141\\
9&01:32:42.93&30:35:38.6&17.61&0.32&\nodata&8.56&4.63&Cluster&CS U106; Hilt L; MD 1; Ma 2002a; Ma 2004b\\
10&01:32:44.30&30:40:12.4&18.75&\nodata&\nodata&10.28&5.58&Cluster&CBF 161; CS U88; Ma 2002b; Ma 2004a\\
11&01:32:45.31&30:30:24.3&\nodata&\nodata&\nodata&\nodata&\nodata&Unknown&CS U127\\
12&01:32:46.80&30:33:35.3&\nodata&\nodata&\nodata&\nodata&\nodata&Stellar&CS U108\\
13&01:32:51.77&30:33:05.3&\nodata&\nodata&\nodata&\nodata&\nodata&Stellar&CS C34\\
14&01:32:51.78&30:29:47.8&\nodata&\nodata&\nodata&\nodata&\nodata&Stellar&CS U128\\
15&01:32:51.82&30:29:36.4&\nodata&\nodata&\nodata&\nodata&\nodata&Cluster&CS U129\\
16&01:32:52.65&30:14:30.9&18.76&0.46&\nodata&8.01&3.72&Cluster&MD 2; Ma 2002a; Ma 2004b\\
17&01:32:52.70&30:32:00.4&\nodata&\nodata&\nodata&\nodata&\nodata&Unknown&CS U116\\
18&01:32:52.87&30:34:10.2&\nodata&\nodata&\nodata&\nodata&\nodata&Unknown&CS U107\\
19&01:32:53.80&30:37:52.9&\nodata&\nodata&\nodata&\nodata&\nodata&Stellar&CS H26\\
20&01:32:54.31&30:55:29.5&\nodata&\nodata&\nodata&\nodata&\nodata&Unknown&CS C7\\
21&01:32:54.63&30:23:20.6&\nodata&\nodata&\nodata&\nodata&\nodata&Unknown&CS U157\\
22&01:32:54.95&30:46:25.4&\nodata&\nodata&\nodata&\nodata&\nodata&Unknown&CS U55\\
23&01:32:55.47&30:29:22.2&\nodata&\nodata&\nodata&\nodata&\nodata&Cluster&CS H37\\
24&01:32:56.09&30:38:25.7&17.63&0.09&0.63&7.22&3.60&Stellar&MKKSS 1; Ma 2002c\\
25&01:32:56.18&30:25:45.8&\nodata&\nodata&\nodata&\nodata&\nodata&Stellar&CS U149\\
26&01:32:56.32&30:14:58.9&18.06&-0.04&\nodata&10.04&5.88&Cluster&Hilt K; MD 3; Ma 2002a; Ma 2004b\\
27&01:32:56.36&30:44:51.2&\nodata&\nodata&\nodata&\nodata&\nodata&Stellar&CS H16\\
28&01:32:57.60&30:55:42.7&\nodata&\nodata&\nodata&\nodata&\nodata&Cluster&CS C8\\
29&01:32:58.63&30:47:57.3&\nodata&\nodata&\nodata&\nodata&\nodata&Cluster&CS C28\\
30&01:32:59.28&30:23:04.4&\nodata&\nodata&\nodata&\nodata&\nodata&Unknown&CS H47\\
31&01:33:00.37&30:26:47.7&\nodata&\nodata&\nodata&\nodata&\nodata&Unknown&CS U142\\
32&01:33:00.54&30:45:17.6&\nodata&\nodata&\nodata&\nodata&\nodata&Cluster&CS U65\\
33&01:33:00.89&30:25:32.7&\nodata&\nodata&\nodata&\nodata&\nodata&Unknown&CS H44\\
34&01:33:01.10&30:35:45.1&20.40&-0.14&\nodata&\nodata&\nodata&Cluster&CBF 39; Ma 2001\\
35&01:33:02.40&30:34:44.5&18.50&-0.34&\nodata&6.42&\nodata&Cluster&CBF 40; Ma 2001\\
36&01:33:04.91&30:25:27.0&\nodata&\nodata&\nodata&\nodata&\nodata&Unknown&CS U151\\
37&01:33:05.56&30:36:40.4&\nodata&\nodata&\nodata&\nodata&\nodata&Unknown&CS U96\\
38&01:33:06.40&30:37:35.8&19.78&2.53&2.03&10.30&5.09&Unknown&MKKSS 2; Ma 2002c\\
39&01:33:07.37&30:23:14.3&\nodata&\nodata&\nodata&\nodata&\nodata&Stellar&CS H46\\
40&01:33:08.11&30:28:00.2&19.11&0.85&1.01&9.15&4.42&Cluster&CBF 86; CS U140; Ma 2002b\\
41&01:33:09.82&30:12:50.7&18.68&0.19&\nodata&7.96&3.71&Cluster&Hilt V; MD 4; Ma 2002a; Ma 2004b\\
42&01:33:10.11&30:29:56.9&18.54&\nodata&0.47&6.58&2.97&Cluster&CBF 89; Ma 2002b\\
43&01:33:11.61&30:13:14.1&18.75&0.27&\nodata&10.22&5.52&Galaxy&Hilt U; MD 5; Ma 2002a; Ma 2004b\\
44&01:33:13.80&30:29:03.6&19.04&\nodata&1.15&10.06&5.32&Cluster&CBF 87; Ma 2002b\\
45&01:33:13.87&30:29:05.1&19.56&0.44&\nodata&8.96&4.15&Cluster&CBF 53; Ma 2001\\
46&01:33:13.88&30:28:24.4&18.27&\nodata&1.10&\nodata&\nodata&Cluster&CBF 85\\
47&01:33:13.90&30:29:44.7&18.20&\nodata&0.00&6.48&\nodata&Cluster&CBF 88; Ma 2002b\\
48&01:33:14.29&30:27:11.1&18.31&0.42&\nodata&6.96&3.01&Unknown&CS U138; MD 7; Ma 2002a; Ma 2004b\\
49&01:33:14.30&30:28:22.8&18.35&1.09&\nodata&10.27&5.68&Cluster&CBF 54; CS U137; Hilt S; MD 8; Ma 2002a; Ma 2004b; Ma 2001; Ma 2004a\\
50&01:33:14.61&30:51:37.8&\nodata&\nodata&\nodata&\nodata&\nodata&Galaxy&CS H7\\
51&01:33:15.09&30:54:12.7&\nodata&\nodata&\nodata&\nodata&\nodata&Unknown&CS H5\\
52&01:33:15.17&30:32:53.0&19.05&\nodata&0.62&9.11&4.33&Cluster&CBF 144; Ma 2002b\\
53&01:33:16.10&30:20:56.7&18.28&0.19&0.61&6.96&3.02&Cluster&MD 6; CS M15; Ma 2002a; Ma 2004b\\
54&01:33:16.63&30:34:35.7&19.33&\nodata&0.62&8.01&3.50&Cluster&CBF 145; Ma 2002b\\
55&01:33:18.20&30:43:48.1&\nodata&\nodata&\nodata&\nodata&\nodata&Unknown&CS U66\\
56&01:33:18.87&30:26:45.0&\nodata&\nodata&\nodata&\nodata&\nodata&Unknown&CS H40\\
57&01:33:19.21&30:23:22.5&\nodata&\nodata&\nodata&\nodata&\nodata&Galaxy&CS U156\\
58&01:33:19.41&30:48:48.7&\nodata&\nodata&\nodata&\nodata&\nodata&Unknown&CS U37\\
59&01:33:20.40&30:40:23.3&17.17&0.24&0.79&8.41&4.66&Unknown&MKKSS 3; Ma 2002c\\
60&01:33:20.48&30:26:15.2&\nodata&\nodata&\nodata&\nodata&\nodata&Stellar&CS H41\\
61&01:33:21.18&30:37:55.5&\nodata&\nodata&\nodata&\nodata&\nodata&Unknown&CS H25\\
62&01:33:21.57&30:31:51.4&\nodata&\nodata&\nodata&\nodata&\nodata&Stellar&CS H33\\
63&01:33:21.66&30:37:48.4&\nodata&\nodata&\nodata&\nodata&\nodata&Unknown&CS U95\\
64&01:33:21.90&31:01:11.2&17.07&1.15&\nodata&10.00&5.96&Galaxy&MD 18; Ma 2002a; Ma 2004b\\
65&01:33:22.10&30:45:34.3&19.19&0.66&0.88&\nodata&\nodata&Cluster&CS C31\\
66&01:33:22.11&30:40:28.4&18.65&\nodata&\nodata&9.11&4.73&Cluster&CBF 59; Ma 2001; Ma 2004a\\
67&01:33:22.16&30:40:26.0&18.29&\nodata&0.38&6.94&3.04&Cluster&CBF 95; Ma 2002b\\
68&01:33:22.32&30:40:59.4&18.58&0.10&0.57&8.36&4.08&Cluster&CBF 60; CBF 94; Ma 2001; Ma 2002b\\
69&01:33:22.38&30:30:14.3&\nodata&\nodata&\nodata&\nodata&\nodata&Stellar&CS H36\\
70&01:33:23.10&30:33:00.5&17.38&0.31&0.38&7.86&4.29&Cluster&MKKSS 4; CS H32; MD 10; Ma 2002a; Ma 2004b; Ma 2002c\\
71&01:33:23.11&30:32:22.9&\nodata&\nodata&\nodata&\nodata&\nodata&Unknown&CS U115\\
72&01:33:23.30&30:46:09.0&19.36&0.61&0.60&\nodata&\nodata&Unknown&CS U50\\
73&01:33:23.44&30:22:31.0&16.66&1.18&\nodata&9.78&6.02&Galaxy&CS C37; MD 9; Ma 2002a; Ma 2004b\\
74&01:33:23.90&30:40:26.0&19.07&\nodata&0.81&6.92&2.72&Cluster&CBF 96; Ma 2002b\\
75&01:33:24.61&30:32:56.1&19.90&0.20&\nodata&\nodata&\nodata&Cluster&CBF 17; Ma 2001\\
76&01:33:24.85&30:33:55.0&19.87&0.01&\nodata&7.28&2.84&Cluster&CBF 18; Ma 2001\\
77&01:33:25.60&30:29:56.8&18.43&0.76&1.43&10.24&5.64&Cluster&MKKSS 5; CS U126; Ma 2002c\\
78&01:33:25.60&30:45:30.5&18.82&0.40&\nodata&\nodata&\nodata&Cluster&CS U64\\
79&01:33:25.65&30:27:52.8&18.67&\nodata&\nodata&\nodata&\nodata&Unknown&CS U136\\
80&01:33:26.00&30:36:24.3&17.87&0.29&0.79&8.51&4.44&Unknown&MKKSS 6; Ma 2002c\\
81&01:33:26.37&30:41:06.9&18.61&\nodata&0.46&7.70&3.57&Cluster&CBF 92; Ma 2002b\\
82&01:33:26.47&30:55:10.9&\nodata&\nodata&\nodata&\nodata&\nodata&Unknown&CS U6\\
83&01:33:26.49&30:41:11.6&19.11&\nodata&0.55&6.60&2.62&Cluster&CBF 93; Ma 2002b\\
84&01:33:26.53&30:27:00.4&18.76&0.36&\nodata&\nodata&\nodata&Unknown&CS U143\\
85&01:33:26.75&30:33:21.4&17.45&0.27&\nodata&7.08&3.62&Cluster&CBF 16; MD 12; CS M10; Ma 2001\\
86&01:33:26.94&30:34:52.6&18.75&0.21&\nodata&6.94&2.98&Cluster&CBF 19; Ma 2001\\
87&01:33:27.40&30:41:59.8&18.28&\nodata&1.01&10.28&5.69&Cluster&CBF 97; Ma 2002b; Ma 2004a\\
88&01:33:27.96&30:37:28.4&18.86&\nodata&0.78&6.86&2.85&Cluster&CBF 124; Ma 2002b\\
89&01:33:27.98&30:32:43.1&18.45&0.20&0.43&6.72&2.99&Unknown&MKKSS 7; Ma 2002c\\
90&01:33:28.00&30:21:06.2&19.65&\nodata&0.44&6.84&2.91&Cluster&CBF 115; Ma 2002b\\
91&01:33:28.13&30:58:30.6&17.78&0.84&\nodata&10.03&5.70&Cluster&CS C1; MD 21; Ma 2002a; Ma 2004b\\
92&01:33:28.23&30:46:39.9&19.14&0.51&\nodata&\nodata&\nodata&Unknown&CS U51\\
93&01:33:28.40&30:36:23.1&17.35&0.69&0.79&8.76&4.83&Cluster&CBF 151; CS H28; MD 14; Ma 2002b\\
94&01:33:28.70&30:36:37.5&17.38&0.52&0.76&8.01&4.35&Cluster&CBF 150; CS H27; MD 15; Ma 2002b\\
95&01:33:28.72&30:41:35.0&17.17&0.76&1.03&\nodata&\nodata&Cluster&CBF 90; MD 17;CS U77; Ma 2004a\\
96&01:33:29.30&30:44:01.6&19.12&0.81&0.73&\nodata&\nodata&Unknown&CS U67\\
97&01:33:29.40&30:23:59.2&\nodata&\nodata&\nodata&\nodata&\nodata&Unknown&CS U152\\
98&01:33:29.48&30:30:02.1&18.16&0.21&0.65&8.01&3.99&Unknown&MKKSS 8; Ma 2002c\\
99&01:33:29.85&30:26:49.1&18.64&0.68&\nodata&\nodata&\nodata&Stellar&CS H42\\
100&01:33:30.07&30:49:29.0&20.01&\nodata&0.60&\nodata&\nodata&Cluster&CBF 111\\
101&01:33:30.68&30:26:31.8&18.01&1.13&\nodata&10.05&5.67&Cluster&CS C36; MD 13; Ma 2002a; Ma 2004b\\
102&01:33:30.70&30:22:21.4&18.18&0.76&0.89&9.28&4.94&Cluster&CBF 114; CS C38; Hilt T; MD 11; Ma 2004a\\
103&01:33:30.90&30:49:11.8&18.52&0.65&0.77&8.96&4.52&Cluster&SBGHS 7; CBF 110; CS C24; MD 20; Ma 2002b\\
104&01:33:30.92&30:37:12.9&18.22&\nodata&0.56&\nodata&\nodata&Cluster&CBF 123\\
105&01:33:31.00&30:36:52.6&18.21&0.32&1.28&6.96&3.02&Cluster&CBF 125; MKKSS 9; Ma 2002b; Ma 2004a; Ma 2002c\\
106&01:33:31.10&30:33:45.5&18.67&0.21&\nodata&7.81&3.57&Cluster&CBF 13; CS U110; Ma 2001\\
107&01:33:31.22&30:33:33.5&18.12&0.20&\nodata&7.74&3.78&Unknown&CS U109; MD 16; Ma 2002a; Ma 2004b\\
108&01:33:31.22&30:54:41.6&18.72&0.78&\nodata&\nodata&\nodata&Chip Gap&CS C9\\
109&01:33:31.25&30:50:07.0&19.27&\nodata&0.43&7.81&3.44&Cluster&CBF 113; CS U36; Ma 2002b\\
110&01:33:31.39&30:40:20.4&17.91&0.58&0.92&9.26&5.03&Cluster&CBF 22; CBF 91; CS U87; Ma 2001; Ma 2002b; Ma 2004a\\
111&01:33:31.86&30:54:40.1&19.74&0.61&\nodata&\nodata&\nodata&Chip Gap&CS C10\\
112&01:33:32.00&30:46:25.0&18.80&0.57&0.84&\nodata&\nodata&Unknown&CS U52\\
113&01:33:32.01&30:33:21.8&17.34&0.17&\nodata&6.98&3.54&Cluster&CBF 14; Ma 2001\\
114&01:33:32.17&30:40:31.9&18.90&-0.04&\nodata&6.52&2.83&Cluster&CBF 24; Ma 2001\\
115&01:33:32.19&30:56:04.9&19.57&0.71&\nodata&\nodata&\nodata&Cluster&CS C6\\
116&01:33:32.36&30:38:28.0&18.52&\nodata&0.89&9.95&5.25&Cluster&CBF 126; Ma 2002b\\
117&01:33:32.43&30:38:24.5&18.30&\nodata&\nodata&9.54&5.09&Cluster&CBF 20; Ma 2001; Ma 2004a\\
118&01:33:32.59&30:39:24.5&18.84&0.69&\nodata&8.01&3.70&Cluster&CBF 25; Ma 2001\\
119&01:33:32.72&30:36:55.2&17.71&\nodata&\nodata&6.84&3.03&Cluster&CBF 148; Ma 2002b\\
120&01:33:32.75&30:31:45.1&19.19&0.40&\nodata&\nodata&\nodata&Cluster&CBF 42; Ma 2001\\
121&01:33:33.00&30:49:41.7&18.65&0.29&0.90&9.21&4.69&Cluster&SBGHS 8; CBF 112; CS C23; Hilt P; MD 22; Ma 200b; Ma 2004a\\
122&01:33:33.28&30:48:30.5&18.38&0.57&0.82&8.86&4.51&Cluster&SBGHS 5; CBF 107; CS C25; Ma 2002b\\
123&01:33:33.57&30:36:35.8&18.69&\nodata&\nodata&\nodata&\nodata&Cluster&CBF 149\\
124&01:33:33.72&30:40:03.0&18.27&0.57&\nodata&8.96&4.62&Cluster&CBF 23; CS U86; Ma 2001\\
125&01:33:34.17&30:44:00.2&\nodata&\nodata&\nodata&\nodata&\nodata&Unknown&CS U68\\
126&01:33:34.38&30:42:01.3&18.64&0.24&\nodata&6.94&2.99&Cluster&CBF 37; Ma 2001\\
127&01:33:34.68&30:48:21.2&19.14&\nodata&1.01&8.51&3.92&Cluster&SBGHS 6; CBF 108; Ma 2002b\\
128&01:33:34.90&30:37:05.6&\nodata&\nodata&\nodata&\nodata&\nodata&Unknown&CS H29\\
129&01:33:34.96&30:55:06.1&20.52&0.85&\nodata&\nodata&\nodata&Unknown&CS U7\\
130&01:33:35.10&30:49:00.0&18.30&1.07&1.20&9.90&5.42&Cluster&SBGHS 4; CBF 106; CS H10; Ma 2004a\\
131&01:33:35.27&30:33:11.6&19.04&0.48&\nodata&8.01&3.66&Cluster&CBF 15; CBF 45; Ma 2001\\
132&01:33:35.62&30:38:36.7&17.71&-0.19&\nodata&6.58&3.13&Cluster&CBF 21; Ma 2001\\
133&01:33:35.94&30:36:28.8&\nodata&\nodata&\nodata&\nodata&\nodata&Stellar&CS U97\\
134&01:33:36.19&30:47:55.1&18.17&\nodata&0.46&8.66&4.40&Cluster&SBGHS 2; CBF 109; Ma 2002b\\
135&01:33:36.30&30:56:15.9&\nodata&\nodata&\nodata&\nodata&\nodata&Stellar&CS U5\\
136&01:33:36.70&30:27:08.0&\nodata&\nodata&\nodata&\nodata&\nodata&Unknown&CS U144\\
137&01:33:36.77&30:43:23.1&\nodata&\nodata&\nodata&\nodata&\nodata&Unknown&CS H18\\
138&01:33:36.79&30:49:17.5&18.63&\nodata&0.82&\nodata&\nodata&Cluster&SBGHS 9\\
139&01:33:36.98&30:37:12.0&18.58&0.05&0.11&8.01&3.79&Unknown&MKKSS 10; Ma 2002c\\
140&01:33:37.24&30:34:13.9&17.15&0.08&0.39&6.94&3.57&Cluster&MKKSS 11; CS U111; Ma 2002c\\
141&01:33:37.60&30:28:04.6&17.71&0.14&0.94&8.36&4.46&Unknown&CS U134; Hilt I; MD 19; Ma 2002a; Ma 2004b\\
142&01:33:37.81&30:50:32.3&19.21&\nodata&1.08&\nodata&\nodata&Cluster&SBGHS 10\\
143&01:33:38.00&30:38:02.2&17.49&0.69&1.22&9.63&5.47&Unknown&MKKSS 12; Ma 2004a; Ma 2004a; Ma 2002c\\
144&01:33:38.04&30:33:05.4&17.72&0.29&0.39&7.22&3.53&Cluster&CBF 44; MKKSS 13; Ma 2001; Ma 2002c\\
145&01:33:38.08&30:33:17.6&18.47&0.76&\nodata&8.96&4.53&Cluster&CBF 46; Ma 2001\\
146&01:33:38.14&30:42:22.9&18.66&0.20&\nodata&8.26&3.92&Cluster&CBF 38; Ma 2001\\
147&01:33:38.19&30:43:23.9&\nodata&\nodata&\nodata&\nodata&\nodata&Unknown&CS U69\\
148&01:33:38.63&30:46:10.7&\nodata&\nodata&\nodata&\nodata&\nodata&Unknown&CS U53\\
149&01:33:39.46&30:55:59.7&\nodata&\nodata&\nodata&\nodata&\nodata&Cluster&CS C5\\
150&01:33:39.46&30:56:18.0&\nodata&\nodata&\nodata&\nodata&\nodata&Unknown&CS U4\\
151&01:33:39.49&30:48:48.2&18.54&0.48&0.63&\nodata&\nodata&Cluster&SBGHS 3; CS C26\\
152&01:33:39.69&30:31:09.2&16.44&0.16&-0.04&7.46&4.27&Unknown&MKKSS 14; Ma 2002c\\
153&01:33:39.71&30:32:29.2&18.57&0.76&\nodata&7.12&3.16&Cluster&CBF 43; Ma 2001\\
154&01:33:39.94&30:38:26.2&15.90&0.30&0.89&7.00&4.05&Unknown&MKKSS 15; Ma 2002c\\
155&01:33:40.08&30:21:37.2&19.61&0.71&\nodata&\nodata&\nodata&Cluster&CBF 41; Ma 2001\\
156&01:33:40.38&30:43:58.0&17.20&0.14&0.57&7.22&3.85&Unknown&MKKSS 16; Ma 2002c\\
157&01:33:40.56&30:49:04.7&19.43&0.36&\nodata&\nodata&\nodata&Unknown&CS U38\\
158&01:33:41.14&30:25:50.4&\nodata&\nodata&\nodata&\nodata&\nodata&Unknown&CS H43\\
159&01:33:41.20&30:29:53.8&19.15&0.69&0.80&\nodata&\nodata&Unknown&CS U125\\
160&01:33:41.32&30:52:56.8&\nodata&\nodata&\nodata&\nodata&\nodata&Stellar&CS H6\\
161&01:33:41.54&30:42:44.9&\nodata&\nodata&\nodata&\nodata&\nodata&Unknown&CS U76\\
162&01:33:41.56&30:30:24.1&\nodata&\nodata&\nodata&\nodata&\nodata&Unknown&CS U124\\
163&01:33:41.60&30:28:09.2&\nodata&\nodata&\nodata&\nodata&\nodata&Unknown&CS U135\\
164&01:33:41.60&30:41:43.4&17.08&0.38&1.03&8.51&4.77&Stellar&MKKSS 17; Ma 2002c\\
165&01:33:41.60&30:48:08.5&18.78&\nodata&0.51&\nodata&\nodata&Cluster&SBGHS 1\\
166&01:33:41.94&30:49:20.1&19.43&\nodata&1.43&\nodata&\nodata&Cluster&SBGHS 12\\
167&01:33:42.00&30:26:53.5&19.83&0.52&0.83&\nodata&\nodata&Unknown&CS U133\\
168&01:33:42.71&30:43:49.6&18.13&0.36&\nodata&8.06&3.99&Cluster&CBF 50; Ma 2001\\
169&01:33:42.96&30:42:53.0&17.16&0.09&-0.09&6.82&3.39&Unknown&MKKSS 18; Ma 2002c\\
170&01:33:43.02&30:44:40.8&\nodata&\nodata&\nodata&\nodata&\nodata&Stellar&CS H17\\
171&01:33:43.80&30:40:56.7&18.59&0.03&\nodata&7.26&3.31&Cluster&CBF 1; Ma 2001\\
172&01:33:43.85&30:32:10.4&17.55&0.16&-0.06&8.06&4.33&Cluster&CBF 47; MKKSS 19; Ma 2001; Ma 2002c\\
173&01:33:44.10&30:26:50.2&18.77&0.48&0.58&\nodata&\nodata&Unknown&CS U132\\
174&01:33:44.10&30:30:00.8&18.26&0.40&0.58&\nodata&\nodata&Unknown&CS H35\\
175&01:33:44.15&30:48:36.0&19.10&\nodata&\nodata&\nodata&\nodata&Unknown&CS U39\\
176&01:33:44.51&30:37:52.7&17.52&-0.10&-1.06&8.26&4.47&Unknown&MKKSS 20; CS U94; Ma 2002c\\
177&01:33:44.66&30:21:09.4&\nodata&\nodata&\nodata&\nodata&\nodata&Stellar&CS H49\\
178&01:33:45.10&30:47:46.7&16.21&0.78&1.12&9.60&5.99&Cluster&CBF 61; MKKSS 21; CS U49; Hilt E; MD 24; Ma 2004a; Ma 2002c\\
179&01:33:45.14&30:49:09.2&19.00&\nodata&0.61&8.81&4.20&Cluster&SBGHS 11; CBF 64; Ma 2002b\\
180&01:33:45.50&30:30:40.7&18.98&0.70&1.12&\nodata&\nodata&Cluster&CS U123\\
181&01:33:45.80&30:27:17.3&18.38&0.51&0.61&7.18&3.22&Cluster&CBF 139; CS U131; Ma 2002b\\
182&01:33:46.29&30:47:51.0&18.77&\nodata&0.94&9.28&4.57&Cluster&CBF 62; Ma 2002b\\
183&01:33:47.00&30:45:58.8&18.86&0.67&0.18&\nodata&\nodata&Unknown&CS U54\\
184&01:33:47.00&30:46:36.3&\nodata&\nodata&\nodata&9.76&\nodata&Cluster&CBF 69; Ma 2002b; Ma 2004a\\
185&01:33:48.07&30:54:51.7&\nodata&\nodata&\nodata&\nodata&\nodata&Chip Gap&CS C11\\
186&01:33:48.46&30:45:38.7&18.62&0.81&\nodata&9.11&4.56&Cluster&CBF 52; CS U56; Ma 2001\\
187&01:33:48.65&30:47:42.6&19.42&\nodata&0.82&\nodata&\nodata&Cluster&CBF 65\\
188&01:33:48.71&30:24:17.0&\nodata&\nodata&\nodata&\nodata&\nodata&Chip Gap&CS U150; CS U153\\
189&01:33:49.15&30:49:01.5&19.31&\nodata&0.55&\nodata&\nodata&Cluster&CBF 63\\
190&01:33:49.36&30:47:12.5&17.96&\nodata&0.60&7.02&3.16&Cluster&CBF 68; Ma 2002b\\
191&01:33:49.62&30:34:25.7&18.01&\nodata&0.80&8.06&4.06&Cluster&CBF 137; Ma 2002b\\
192&01:33:50.19&30:34:18.8&18.81&\nodata&0.85&8.81&4.26&Cluster&CBF 136; Ma 2002b\\
193&01:33:50.27&30:31:11.0&18.49&\nodata&0.92&\nodata&\nodata&Cluster&CBF 138\\
194&01:33:50.70&30:58:50.3&17.58&0.04&\nodata&7.28&3.69&Cluster&Hilt J; MD 28; Ma 2004b\\
195&01:33:50.73&30:44:56.2&18.84&0.46&\nodata&7.81&3.61&Cluster&CBF 51; Ma 2001\\
196&01:33:50.85&30:28:59.9&19.27&\nodata&0.52&\nodata&\nodata&Cluster&CBF 105\\
197&01:33:50.85&30:38:34.5&16.39&\nodata&0.08&7.22&4.26&Cluster&CBF 127; Ma 2002b\\
198&01:33:50.90&30:38:55.5&16.76&0.21&\nodata&6.86&3.70&Cluster&CBF 2; Ma 2001\\
199&01:33:50.90&30:31:44.8&17.85&0.13&\nodata&8.01&4.14&Cluster&CBF 4; CS U117; MD 26; Ma 2001\\
200&01:33:50.90&30:34:37.0&\nodata&\nodata&\nodata&\nodata&\nodata&Unknown&CS U112\\
201&01:33:51.24&30:34:13.2&18.11&\nodata&0.59&6.96&3.02&Cluster&CBF 133; Ma 2002b\\
202&01:33:51.30&30:50:55.8&18.61&0.79&0.50&\nodata&\nodata&Unknown&CS C21\\
203&01:33:51.31&30:34:37.0&\nodata&\nodata&\nodata&\nodata&\nodata&Unknown&CS U113\\
204&01:33:51.80&30:31:47.2&18.62&0.07&\nodata&8.01&3.76&Cluster&CBF 5; Ma 2001\\
205&01:33:52.10&30:47:16.2&18.83&\nodata&0.29&\nodata&\nodata&Unknown&CS U48\\
206&01:33:52.20&30:29:03.8&17.29&0.86&1.12&9.70&5.59&Cluster&CBF 104; MKKSS 22; CS H38; Hilt H; MD 25; Ma 2004a; Ma 2002c\\
207&01:33:52.38&30:35:00.8&18.68&\nodata&0.53&6.98&2.98&Cluster&CBF 132; Ma 2002b\\
208&01:33:52.39&30:34:21.1&18.68&\nodata&0.84&\nodata&\nodata&Cluster&CBF 134\\
209&01:33:52.40&30:50:17.1&19.20&0.31&0.70&\nodata&\nodata&Unknown&CS U22\\
210&01:33:52.67&30:48:10.1&19.41&\nodata&0.47&\nodata&\nodata&Cluster&CBF 66\\
211&01:33:53.14&30:48:33.8&\nodata&\nodata&\nodata&\nodata&\nodata&Unknown&CS U40\\
212&01:33:53.40&30:33:02.8&18.86&0.35&\nodata&6.62&2.84&Cluster&CBF 8; Ma 2001\\
213&01:33:53.43&30:57:18.0&\nodata&\nodata&\nodata&\nodata&\nodata&Unknown&CS H2\\
214&01:33:53.69&30:48:21.5&17.45&\nodata&0.42&7.72&4.13&Cluster&CBF 67; Ma 2002b\\
215&01:33:54.10&30:33:09.7&17.33&-0.07&0.55&7.32&3.89&Cluster&CBF 7; MKKSS 23; Ma 2001; Ma 2002c\\
216&01:33:54.38&30:21:51.9&18.54&0.68&\nodata&9.01&4.56&Cluster&CS U158; MD 23; Ma 2002a; Ma 2004b\\
217&01:33:54.63&30:34:48.3&18.83&\nodata&1.11&6.94&2.86&Cluster&CBF 135; Ma 2002b\\
218&01:33:54.73&30:48:43.7&\nodata&\nodata&\nodata&\nodata&\nodata&Cluster&BEA 24; CS U41\\
219&01:33:54.75&30:45:28.4&18.12&0.27&\nodata&6.62&3.00&Cluster&CBF 30; Ma 2001\\
220&01:33:54.80&30:32:15.8&17.86&-0.19&\nodata&6.82&3.07&Cluster&CBF 6; Ma 2001\\
221&01:33:55.00&30:32:14.5&17.42&0.29&\nodata&6.62&3.15&Cluster&CBF 3; Ma 2001\\
222&01:33:55.18&30:47:58.0&16.64&0.28&0.60&7.70&4.44&Cluster&MKKSS 24; Hilt D; MD 27; CS M4; Ma 2002a; Ma 2004b; Ma 2002c\\
223&01:33:55.32&30:37:34.0&\nodata&\nodata&\nodata&\nodata&\nodata&Unknown&CS U93\\
224&01:33:55.35&30:45:43.5&\nodata&\nodata&\nodata&\nodata&\nodata&Unknown&CS U63\\
225&01:33:55.50&30:57:06.3&\nodata&\nodata&\nodata&\nodata&\nodata&Stellar&CS U2\\
226&01:33:55.87&30:29:34.3&18.22&1.06&\nodata&\nodata&\nodata&Galaxy&CS M12\\
227&01:33:55.90&30:52:28.7&19.10&0.31&0.18&\nodata&\nodata&Unknown&CS U10\\
228&01:33:56.18&30:38:39.8&17.38&\nodata&0.92&6.94&3.26&Cluster&CBF 129; Ma 2002b\\
229&01:33:56.21&30:45:51.8&18.58&\nodata&0.63&8.01&3.85&Cluster&CBF 146; Ma 2002b\\
230&01:33:56.41&30:29:28.4&18.16&1.59&\nodata&\nodata&\nodata&Galaxy&CS M11\\
231&01:33:56.50&30:36:10.6&18.26&\nodata&0.90&9.30&5.04&Cluster&CBF 131; Ma 2002b\\
232&01:33:56.93&30:41:38.4&18.79&\nodata&1.16&\nodata&\nodata&Cluster&CBF 102\\
233&01:33:56.93&30:49:26.8&\nodata&\nodata&\nodata&\nodata&\nodata&Cluster&BEA 18\\
234&01:33:57.10&30:50:31.5&18.72&0.95&1.14&\nodata&\nodata&Cluster&CS U23\\
235&01:33:57.10&30:48:03.5&\nodata&\nodata&\nodata&\nodata&\nodata&Cluster&BEA 23\\
236&01:33:57.16&30:40:20.7&18.75&\nodata&0.61&\nodata&\nodata&Cluster&CBF 140\\
237&01:33:57.28&30:39:15.3&17.84&\nodata&0.65&9.11&4.84&Cluster&CBF 128; Ma 2002b\\
238&01:33:57.35&30:41:28.5&18.52&\nodata&0.75&6.62&2.93&Cluster&CBF 103; Ma 2002b\\
239&01:33:57.40&30:52:17.9&18.02&0.25&0.54&8.01&4.07&Cluster&CS C17; Hilt W; MD 31; Ma 2002a; Ma 2004b\\
240&01:33:57.66&30:41:32.6&18.72&\nodata&0.74&8.06&3.74&Cluster&CBF 101; Ma 2002b\\
241&01:33:57.84&30:35:31.8&18.19&0.64&0.46&9.34&4.97&Cluster&CBF 49; MKKSS 25; Ma 2001; Ma 2004a Ma 2002c\\
242&01:33:57.85&30:49:04.9&\nodata&\nodata&\nodata&\nodata&\nodata&Cluster&BEA 20\\
243&01:33:57.87&30:33:25.7&17.00&\nodata&\nodata&7.18&3.92&Cluster&CBF 159; Ma 2002b\\
244&01:33:57.93&31:04:08.7&19.39&\nodata&\nodata&\nodata&\nodata&Unknown&Hilt O\\
245&01:33:58.01&30:45:45.2&17.14&0.27&0.02&8.01&4.49&Cluster&CBF 33; MKKSS 26; CS H14; MD 30; Ma 2001; Ma 2002c\\
246&01:33:58.03&30:39:26.2&17.54&0.17&\nodata&7.24&3.70&Cluster&CBF 26; Ma 2001\\
247&01:33:58.10&30:38:15.5&17.84&\nodata&0.67&9.06&4.94&Cluster&CBF 130; Ma 2002b\\
248&01:33:58.41&30:39:14.9&18.08&0.27&\nodata&8.86&4.60&Cluster&CBF 27; Ma 2001\\
249&01:33:58.58&30:48:42.7&\nodata&\nodata&\nodata&\nodata&\nodata&Cluster&BEA 21\\
250&01:33:58.86&30:34:43.2&18.79&0.42&\nodata&8.01&3.72&Cluster&CBF 48; Ma 2001\\
251&01:33:58.90&30:49:11.0&\nodata&\nodata&\nodata&\nodata&\nodata&Cluster&BEA 19\\
252&01:33:59.07&30:50:05.9&\nodata&\nodata&\nodata&\nodata&\nodata&Cluster&BEA 31\\
253&01:33:59.46&30:48:26.7&\nodata&\nodata&\nodata&\nodata&\nodata&Cluster&BEA 32\\
254&01:33:59.52&30:47:29.5&\nodata&\nodata&\nodata&\nodata&\nodata&Cluster&BEA 28\\
255&01:33:59.52&30:45:49.9&16.70&0.08&\nodata&6.88&3.74&Cluster&CBF 32; Ma 2001\\
256&01:33:59.67&30:47:38.2&\nodata&\nodata&\nodata&\nodata&\nodata&Cluster&BEA 29\\
257&01:33:59.74&30:41:24.4&17.70&\nodata&0.45&6.68&3.19&Cluster&CBF 100; Ma 2002b\\
258&01:33:59.84&30:39:45.4&18.44&0.98&\nodata&8.01&3.89&Cluster&CBF 29; Ma 2001\\
259&01:33:59.92&30:32:44.2&\nodata&\nodata&\nodata&\nodata&\nodata&Unknown&CS U114\\
260&01:34:00.01&30:33:54.3&16.19&\nodata&\nodata&6.98&4.01&Cluster&CBF 158; Ma 2002b\\
261&01:34:00.21&30:37:47.2&16.14&0.64&0.74&8.96&5.48&Unknown&MKKSS 27; Ma 2002c\\
262&01:34:00.27&30:48:36.6&\nodata&\nodata&\nodata&\nodata&\nodata&Cluster&BEA 22; CS U42\\
263&01:34:00.44&30:51:01.2&\nodata&\nodata&\nodata&\nodata&\nodata&Unknown&CS U24\\
264&01:34:00.47&30:41:23.1&18.15&\nodata&0.97&9.11&4.76&Cluster&CBF 99; Ma 2002b\\
265&01:34:00.76&30:50:09.1&\nodata&\nodata&\nodata&\nodata&\nodata&Cluster&BEA 26\\
266&01:34:01.03&30:46:58.8&\nodata&\nodata&\nodata&\nodata&\nodata&Cluster&BEA 30\\
267&01:34:01.31&30:39:23.5&18.17&\nodata&0.76&6.60&3.00&Cluster&CBF 120; Ma 2002b\\
268&01:34:01.56&30:42:31.0&\nodata&\nodata&\nodata&\nodata&\nodata&Cluster&CS U74\\
269&01:34:01.60&30:42:31.1&17.92&0.24&0.08&6.62&3.02&Unknown&MKKSS 28; CS U75; Ma 2002c\\
270&01:34:01.68&30:49:43.9&\nodata&\nodata&\nodata&\nodata&\nodata&Cluster&BEA 25; CS H9\\
271&01:34:01.75&30:32:25.7&18.46&\nodata&\nodata&8.51&4.18&Cluster&CBF 160; Ma 2002b\\
272&01:34:01.99&30:38:10.9&18.43&\nodata&0.48&6.62&2.98&Cluster&CBF 121; Ma 2002b\\
273&01:34:01.99&30:39:37.8&16.35&0.82&1.08&9.80&6.11&Cluster&CBF 28; MKKSS 29; Ma 2001; Ma 2004a; Ma 2002c\\
274&01:34:02.33&30:50:27.8&\nodata&\nodata&\nodata&\nodata&\nodata&Cluster&BEA 14\\
275&01:34:02.48&30:40:40.7&16.52&1.04&1.41&9.11&5.42&Cluster&CBF 98; MKKSS 31; Ma 2002c\\
276&01:34:02.48&30:38:41.1&17.14&0.09&-0.23&7.18&3.85&Unknown&MKKSS 30; Ma 2002c\\
277&01:34:02.59&30:58:10.3&\nodata&\nodata&\nodata&\nodata&\nodata&Galaxy&CS U1\\
278&01:34:02.63&30:49:38.6&17.84&0.28&\nodata&8.36&4.48&Stellar&MD 33; Ma 2002a; Ma 2004b\\
279&01:34:02.77&30:48:36.5&\nodata&\nodata&\nodata&\nodata&\nodata&Cluster&BEA 13\\
280&01:34:02.79&30:46:36.8&17.68&0.29&0.75&7.00&3.30&Unknown&MKKSS 32; Ma 2002c\\
281&01:34:02.90&30:43:20.8&16.40&0.74&1.11&9.21&5.56&Cluster&MKKSS 33; Ma 2004a; Ma 2002c\\
282&01:34:03.09&30:45:35.6&18.51&0.43&\nodata&9.21&4.71&Cluster&CBF 31; Ma 2001\\
283&01:34:03.10&30:42:21.3&\nodata&\nodata&\nodata&\nodata&\nodata&Unknown&CS H20\\
284&01:34:03.12&30:52:13.9&16.83&0.29&\nodata&6.98&3.78&Cluster&MD 37; CS M2; Ma 2002a; Ma 2004b\\
285&01:34:03.12&30:48:11.0&\nodata&\nodata&\nodata&\nodata&\nodata&Cluster&BEA 10\\
286&01:34:03.14&30:46:55.2&\nodata&\nodata&\nodata&\nodata&\nodata&Unknown&CS U47\\
287&01:34:03.34&30:48:28.0&\nodata&\nodata&\nodata&\nodata&\nodata&Cluster&BEA 15\\
288&01:34:03.83&30:29:33.5&18.56&0.46&\nodata&8.51&4.20&Unknown&CS C35; Hilt R; MD 29; Ma 2002a; Ma 2004b\\
289&01:34:03.90&30:47:29.1&17.54&0.58&\nodata&7.06&3.39&Cluster&BEA 7; MD 34; CS M5; Ma 2002a; Ma 2004b\\
290&01:34:04.32&30:39:22.8&17.97&0.10&0.62&7.10&3.20&Unknown&MKKSS 34; Ma 2002c\\
291&01:34:04.47&30:36:56.1&\nodata&\nodata&\nodata&\nodata&\nodata&Unknown&CS U98\\
292&01:34:04.79&30:49:17.9&\nodata&\nodata&\nodata&\nodata&\nodata&Cluster&BEA 27\\
293&01:34:04.80&30:47:39.1&\nodata&\nodata&\nodata&\nodata&\nodata&Stellar&BEA 17\\
294&01:34:05.08&30:49:43.1&\nodata&\nodata&\nodata&\nodata&\nodata&Cluster&BEA 5\\
295&01:34:05.12&30:40:36.7&\nodata&\nodata&\nodata&\nodata&\nodata&Unknown&CS U85\\
296&01:34:05.24&30:57:01.1&\nodata&\nodata&\nodata&\nodata&\nodata&Cluster&CS C2\\
297&01:34:05.40&30:47:50.9&\nodata&\nodata&\nodata&\nodata&\nodata&Stellar&BEA 12\\
298&01:34:05.85&30:49:56.9&\nodata&\nodata&\nodata&\nodata&\nodata&Cluster&BEA 4; CS H8\\
299&01:34:06.20&30:40:12.6&17.05&0.24&0.52&\nodata&\nodata&Unknown&CS U84\\
300&01:34:06.30&30:37:26.1&17.95&\nodata&0.98&9.16&4.97&Cluster&CBF 118; Ma 2002b; Ma 2004a\\
301&01:34:06.40&30:37:30.5&18.25&\nodata&0.94&9.16&4.89&Cluster&CBF 119; Ma 2002b\\
302&01:34:06.59&30:50:18.3&\nodata&\nodata&\nodata&\nodata&\nodata&Cluster&BEA 3\\
303&01:34:06.68&30:48:56.2&\nodata&\nodata&\nodata&\nodata&\nodata&Cluster&BEA 8\\
304&01:34:06.77&30:48:32.8&\nodata&\nodata&\nodata&\nodata&\nodata&Cluster&BEA 9\\
305&01:34:06.79&30:47:27.0&\nodata&\nodata&\nodata&\nodata&\nodata&Cluster&BEA 16\\
306&01:34:06.98&30:32:00.1&\nodata&\nodata&\nodata&\nodata&\nodata&Unknown&CS U118\\
307&01:34:07.02&30:50:57.4&\nodata&\nodata&\nodata&\nodata&\nodata&Cluster&BEA 1\\
308&01:34:07.03&30:49:24.4&\nodata&\nodata&\nodata&\nodata&\nodata&Cluster&BEA 33\\
309&01:34:07.18&30:35:23.1&18.07&0.01&0.08&8.31&4.08&Stellar&MKKSS 35; Ma 2002c\\
310&01:34:07.28&30:38:29.5&18.36&\nodata&0.80&8.01&3.81&Cluster&CBF 117; Ma 2002b\\
311&01:34:07.37&30:47:41.6&\nodata&\nodata&\nodata&\nodata&\nodata&Cluster&BEA 6\\
312&01:34:07.51&30:50:11.1&\nodata&\nodata&\nodata&\nodata&\nodata&Cluster&BEA 2\\
313&01:34:07.73&30:52:18.1&16.84&0.38&\nodata&6.96&3.75&Unknown&MD 40; CS M1; Ma 2002a; Ma 2004b\\
314&01:34:07.78&30:51:41.4&\nodata&\nodata&\nodata&\nodata&\nodata&Stellar&CS U21\\
315&01:34:07.79&30:31:21.2&\nodata&\nodata&\nodata&\nodata&\nodata&Stellar&CS U121\\
316&01:34:08.04&30:38:38.2&16.39&0.95&1.21&10.00&6.27&Cluster&CBF 116; MKKSS 36; Ma 2002c\\
317&01:34:08.08&30:31:18.7&\nodata&\nodata&\nodata&\nodata&\nodata&Unknown&CS U119\\
318&01:34:08.10&30:53:32.1&\nodata&\nodata&\nodata&\nodata&\nodata&Stellar&CS U11\\
319&01:34:08.22&30:34:05.0&\nodata&\nodata&\nodata&\nodata&\nodata&Stellar&CS H31\\
320&01:34:08.53&30:39:02.4&16.27&0.10&0.37&6.66&3.93&Cluster&CBF 141; MKKSS 37; Ma 2002b; Ma 2002c\\
321&01:34:08.63&30:39:22.8&17.34&\nodata&0.65&8.31&4.59&Cluster&CBF 122; Ma 2002b\\
322&01:34:08.70&30:42:55.3&18.79&0.66&0.64&8.36&3.96&Cluster&CBF 152; CS U73; Ma 2002b\\
323&01:34:08.76&30:48:16.1&\nodata&\nodata&\nodata&\nodata&\nodata&Cluster&BEA 11\\
324&01:34:08.77&30:30:57.0&\nodata&\nodata&\nodata&\nodata&\nodata&Stellar&CS H34\\
325&01:34:08.96&30:36:33.8&17.85&0.13&-0.07&6.88&3.14&Unknown&MKKSS 38; CS H30; MD 35; Ma 2002a; Ma 2004b; Ma 2002c\\
326&01:34:09.36&30:47:01.8&\nodata&\nodata&\nodata&\nodata&\nodata&Unknown&CS H13\\
327&01:34:09.71&30:21:30.0&18.42&\nodata&0.51&6.76&2.98&Cluster&CBF 147; CS U159; Ma 2002b\\
328&01:34:09.78&30:52:06.1&\nodata&\nodata&\nodata&\nodata&\nodata&Cluster&CS U20\\
329&01:34:10.09&30:45:29.4&17.48&-0.39&0.01&6.96&3.35&Cluster&MKKSS 39; CS U62; MD 39; Ma 2002a; Ma 2004b; Ma 2002c\\
330&01:34:10.24&30:55:26.6&\nodata&\nodata&\nodata&\nodata&\nodata&Unknown&CS U3\\
331&01:34:10.66&30:45:48.9&16.07&0.12&0.38&7.16&4.20&Unknown&MKKSS 40; Ma 2002c\\
332&01:34:10.67&30:35:16.8&18.31&0.79&\nodata&\nodata&\nodata&Unknown&CS C32\\
333&01:34:11.00&30:40:30.1&17.77&0.31&0.77&8.81&4.77&Cluster&MKKSS 41; CS U83; MD 38; Ma 2002a; Ma 2004b; Ma 2002c\\
334&01:34:11.29&30:24:13.2&\nodata&\nodata&\nodata&\nodata&\nodata&Chip Gap&CS U155\\
335&01:34:11.35&30:41:27.9&18.12&-0.20&\nodata&8.56&4.37&Cluster&MD 41; Ma 2002a; Ma 2004b\\
336&01:34:11.36&30:41:27.9&18.15&0.43&0.80&8.56&4.37&Unknown&MKKSS 42; CS U78; Ma 2002c\\
337&01:34:11.55&30:34:52.5&16.61&0.32&0.67&8.31&4.81&Unknown&MKKSS 43; Ma 2002c\\
338&01:34:11.82&30:42:19.9&18.61&\nodata&\nodata&8.91&4.46&Cluster&CBF 153; Ma 2002b\\
339&01:34:11.86&30:24:10.1&\nodata&\nodata&\nodata&\nodata&\nodata&Chip Gap&CS U154\\
340&01:34:12.68&30:47:05.3&\nodata&\nodata&\nodata&\nodata&\nodata&Stellar&CS H12\\
341&01:34:13.60&30:34:48.5&17.76&0.69&\nodata&\nodata&\nodata&Unknown&CS U105\\
342&01:34:13.69&30:43:18.4&\nodata&\nodata&\nodata&\nodata&\nodata&Unknown&CS U70\\
343&01:34:13.70&30:35:22.2&18.31&0.47&\nodata&\nodata&\nodata&Unknown&CS U104\\
344&01:34:13.80&30:45:31.5&17.77&0.35&\nodata&9.11&4.93&Unknown&CS U61; MD 44; Ma 2002a; Ma 2004b\\
345&01:34:13.84&30:19:47.3&18.31&0.31&\nodata&8.86&4.54&Cluster&MD 32; Ma 2002a; Ma 2004b\\
346&01:34:13.99&30:27:59.0&\nodata&\nodata&\nodata&\nodata&\nodata&Cluster&CS U130\\
347&01:34:14.02&30:39:29.5&18.28&0.47&1.12&8.01&3.95&Cluster&CBF 56; CBF 156; CS U89; Ma 2001; Ma 2002b\\
348&01:34:14.14&30:52:59.8&\nodata&\nodata&\nodata&\nodata&\nodata&Unknown&CS U12\\
349&01:34:14.19&30:36:12.1&18.23&0.55&\nodata&\nodata&\nodata&Unknown&CS U99\\
350&01:34:14.20&30:39:58.4&18.19&0.27&0.73&8.96&4.68&Cluster&CBF 55; MKKSS 44; MD 42;CS U82; Ma 2001; Ma 2002c\\
351&01:34:14.65&30:32:35.0&18.16&0.20&0.59&8.56&4.37&Cluster&MKKSS 45; CS C33; MD 36; Ma 2002a; Ma 2004b; Ma 2002c\\
352&01:34:15.02&30:53:33.9&\nodata&\nodata&\nodata&\nodata&\nodata&Unknown&CS H4\\
353&01:34:15.04&30:41:19.2&17.53&0.35&0.38&6.96&3.34&Cluster&CBF 155; MKKSS 46; CS U79; Ma 2002b; Ma 2002c\\
354&01:34:15.51&30:50:01.7&\nodata&\nodata&\nodata&\nodata&\nodata&Unknown&CS U29\\
355&01:34:15.51&30:42:11.5&17.92&\nodata&\nodata&8.51&4.60&Cluster&CBF 154; CS H19; Ma 2002b\\
356&01:34:15.78&30:27:45.7&\nodata&\nodata&\nodata&\nodata&\nodata&Galaxy&CS U147\\
357&01:34:16.10&30:45:03.9&\nodata&\nodata&\nodata&\nodata&\nodata&Unknown&CS U60\\
358&01:34:16.37&30:47:43.0&\nodata&\nodata&\nodata&\nodata&\nodata&Cluster&CS U46\\
359&01:34:16.38&30:37:49.1&17.72&0.14&\nodata&7.81&4.09&Cluster&CBF 34; Ma 2001\\
360&01:34:16.57&30:40:29.0&18.98&\nodata&\nodata&8.96&4.26&Cluster&CBF 157; Ma 2002b\\
361&01:34:17.54&30:42:36.7&19.67&1.02&\nodata&7.48&3.02&Cluster&CBF 9; Ma 2001\\
362&01:34:17.64&30:28:24.0&\nodata&\nodata&\nodata&\nodata&\nodata&Galaxy&CS U146\\
363&01:34:17.87&30:35:34.1&19.00&\nodata&\nodata&\nodata&\nodata&Stellar&CS U103\\
364&01:34:18.19&30:52:30.0&\nodata&\nodata&\nodata&\nodata&\nodata&Unknown&CS U13\\
365&01:34:18.21&30:53:47.4&\nodata&\nodata&\nodata&\nodata&\nodata&Unknown&CS H3\\
366&01:34:18.26&30:22:02.5&\nodata&\nodata&\nodata&\nodata&\nodata&Unknown&CS H48\\
367&01:34:18.59&30:44:47.8&18.81&\nodata&\nodata&10.30&5.57&Cluster&CBF 11; Ma 2001; Ma 2004a\\
368&01:34:18.69&30:31:37.6&16.75&0.27&0.31&7.38&4.18&Unknown&MKKSS 47; Ma 2002c\\
369&01:34:19.29&30:23:33.0&\nodata&\nodata&\nodata&\nodata&\nodata&Stellar&CS H45\\
370&01:34:19.44&30:46:21.2&16.78&0.26&0.60&8.06&4.59&Unknown&MKKSS 48; Ma 2002c\\
371&01:34:19.89&30:36:12.7&17.16&0.19&0.37&7.81&4.33&Cluster&CBF 35; MKKSS 49; Ma 2001; Ma 2002c\\
372&01:34:20.17&30:39:33.3&18.55&0.48&0.44&6.96&3.02&Cluster&CBF 58; MKKSS 50; CS U91; Ma 2001; Ma 2002c\\
373&01:34:20.78&30:38:33.1&\nodata&\nodata&\nodata&\nodata&\nodata&Unknown&CS H23\\
374&01:34:20.95&30:22:57.6&\nodata&\nodata&\nodata&\nodata&\nodata&Galaxy&CS U145\\
375&01:34:21.43&30:39:40.2&19.38&0.62&\nodata&10.28&5.34&Cluster&CBF 57; Ma 2001\\
376&01:34:21.59&30:36:45.6&18.36&0.42&0.55&7.81&3.87&Cluster&SBGHS 23; CBF 36; Ma 2001\\
377&01:34:21.99&30:44:39.1&18.77&-0.08&\nodata&7.32&3.35&Cluster&CBF 10; CS U59; Ma 2001\\
378&01:34:22.23&30:47:11.0&\nodata&\nodata&\nodata&\nodata&\nodata&Unknown&CS U45\\
379&01:34:22.24&30:30:42.6&\nodata&\nodata&\nodata&\nodata&\nodata&Unknown&CS U120\\
380&01:34:22.81&30:54:26.1&\nodata&\nodata&\nodata&\nodata&\nodata&Chip Gap&CS U9\\
381&01:34:22.92&30:47:33.7&\nodata&\nodata&\nodata&\nodata&\nodata&Stellar&CS H11\\
382&01:34:23.05&30:37:39.8&18.75&\nodata&0.68&\nodata&\nodata&Cluster&SBGHS 22\\
383&01:34:23.13&30:43:46.4&19.01&0.30&\nodata&6.94&2.79&Cluster&CBF 12; Ma 2001\\
384&01:34:23.52&30:25:58.2&17.96&0.16&\nodata&8.01&4.14&Cluster&CS U148; Hilt G; MD 43; Ma 2002a; Ma 2004b\\
385&01:34:24.53&30:53:05.4&18.19&0.80&0.99&10.13&5.65&Cluster&CS C18; MD 49; Ma 2002a; Ma 2004b\\
386&01:34:25.00&30:51:33.3&19.65&0.47&0.06&\nodata&\nodata&Unknown&CS U19\\
387&01:34:25.40&30:41:28.4&17.49&0.66&0.79&9.01&4.99&Cluster&MKKSS 51; CS H21; Hilt F; MD 45; Ma 2002a; Ma 2004b; Ma 2004a; Ma 2002c\\
388&01:34:25.51&30:36:56.8&18.17&0.42&0.85&\nodata&\nodata&Cluster&SBGHS 20; CS U100\\
389&01:34:26.32&30:37:23.3&18.10&\nodata&0.72&\nodata&\nodata&Cluster&SBGHS 21\\
390&01:34:26.39&30:47:13.8&\nodata&\nodata&\nodata&\nodata&\nodata&Unknown&CS U44\\
391&01:34:26.88&30:41:46.2&18.92&0.66&\nodata&\nodata&\nodata&Unknown&CS U80\\
392&01:34:27.10&30:36:42.3&17.69&0.44&0.69&8.91&4.89&Cluster&SBGHS 19; CS U101; MD 46; Ma 2002a; Ma 2004b\\
393&01:34:27.61&30:55:53.3&19.74&\nodata&\nodata&\nodata&\nodata&Cluster&CS C4\\
394&01:34:28.14&30:42:48.2&18.86&0.35&\nodata&\nodata&\nodata&Unknown&CS U72\\
395&01:34:28.19&30:36:17.1&15.95&\nodata&0.46&\nodata&\nodata&Cluster&SBGHS 18; Hilt C\\
396&01:34:28.50&30:37:56.1&18.59&\nodata&0.47&\nodata&\nodata&Cluster&SBGHS 24\\
397&01:34:28.50&30:53:35.9&19.16&0.63&0.63&\nodata&\nodata&Galaxy&CS C15\\
398&01:34:28.71&30:21:44.0&\nodata&\nodata&\nodata&\nodata&\nodata&Unknown&CS U160\\
399&01:34:29.07&30:38:05.4&18.81&\nodata&0.86&8.61&4.10&Cluster&SBGHS 14; CBF 71; Ma 2002b\\
400&01:34:29.10&30:53:20.6&18.35&0.71&0.69&8.81&4.49&Cluster&CS C16; MD 50; Ma 2002a; Ma 2004b\\
401&01:34:29.29&30:56:06.0&18.27&0.87&\nodata&\nodata&\nodata&Cluster&CS C3\\
402&01:34:30.20&30:38:13.0&17.19&0.77&1.01&\nodata&\nodata&Cluster&SBGHS 13; CBF 70; Hilt Q; MD 47; CS M9\\
403&01:34:30.50&30:36:48.2&19.34&0.49&0.46&\nodata&\nodata&Unknown&CS U102\\
404&01:34:31.00&30:57:57.5&\nodata&\nodata&\nodata&\nodata&\nodata&Stellar&CS H1\\
405&01:34:31.07&30:37:41.1&19.50&\nodata&0.09&\nodata&\nodata&Cluster&CBF 80\\
406&01:34:31.74&30:39:14.8&19.43&\nodata&0.57&9.11&4.23&Cluster&CBF 73; Ma 2002b\\
407&01:34:32.29&30:55:09.9&19.77&0.36&\nodata&\nodata&\nodata&Unknown&CS U8\\
408&01:34:32.78&30:54:23.6&\nodata&\nodata&\nodata&\nodata&\nodata&Chip Gap&CS C12\\
409&01:34:32.90&30:38:12.0&18.99&\nodata&1.08&9.11&4.44&Cluster&SBGHS 16; CBF 79; Ma 2002b\\
410&01:34:33.09&30:37:36.3&18.23&\nodata&0.09&6.80&2.99&Cluster&SBGHS 17; CBF 78; Ma 2002b\\
411&01:34:33.12&30:38:14.2&19.46&\nodata&0.58&8.76&4.05&Cluster&SBGHS 15; CBF 76; Ma 2002b\\
412&01:34:33.19&30:38:26.6&19.53&\nodata&1.06&8.76&4.05&Cluster&CBF 75; Ma 2002b\\
413&01:34:33.73&30:39:15.7&18.32&0.30&0.56&9.11&4.76&Cluster&CBF 72; MD 48; CS M8; Ma 2002b\\
414&01:34:34.42&30:42:43.2&\nodata&\nodata&\nodata&\nodata&\nodata&Unknown&CS U71\\
415&01:34:35.16&30:44:59.5&\nodata&\nodata&\nodata&\nodata&\nodata&Stellar&CS H15\\
416&01:34:35.30&30:38:30.1&18.78&\nodata&1.06&9.32&4.72&Cluster&CBF 74; CS H24; Ma 2002b; Ma 2004a\\
417&01:34:36.92&30:03:47.6&\nodata&\nodata&\nodata&\nodata&\nodata&Unknown&CS U139\\
418&01:34:38.39&30:54:49.3&16.83&1.09&\nodata&10.14&6.20&Galaxy&CS C13; Hilt N; MD 53; Ma 2002a; Ma 2004b\\
419&01:34:38.90&30:38:51.8&18.78&\nodata&0.44&9.01&4.60&Cluster&CBF 77; Ma 2002b\\
420&01:34:40.41&30:46:01.3&15.85&\nodata&0.81&6.94&4.01&Cluster&CBF 142; Ma 2002b\\
421&01:34:40.66&30:49:47.3&\nodata&\nodata&\nodata&\nodata&\nodata&Cluster&CS U27\\
422&01:34:40.72&30:53:02.0&19.45&0.42&1.03&6.94&2.65&Cluster&CBF 83; CS U14; Ma 2002b\\
423&01:34:41.20&30:49:52.7&17.90&0.25&0.36&\nodata&\nodata&Stellar&CS U30\\
424&01:34:41.65&30:46:38.8&19.10&0.52&\nodata&\nodata&\nodata&Unknown&CS U57\\
425&01:34:42.80&30:49:19.2&18.93&0.71&0.94&\nodata&\nodata&Cluster&CS U28\\
426&01:34:43.19&30:52:19.1&19.88&\nodata&0.48&\nodata&\nodata&Cluster&CBF 82\\
427&01:34:43.70&30:47:37.9&17.20&0.43&0.72&8.76&4.91&Cluster&CS C27; Hilt B; MD 54; Ma 2002a; Ma 2004b\\
428&01:34:44.20&30:52:18.9&17.64&0.83&1.00&9.95&5.68&Cluster&CBF 81; CS C20; Hilt M; MD 55; Ma 2004a\\
429&01:34:45.09&30:50:33.5&\nodata&\nodata&\nodata&\nodata&\nodata&Cluster&CS U25\\
430&01:34:45.59&30:44:23.0&\nodata&\nodata&\nodata&\nodata&\nodata&Chip Gap&CS C30\\
431&01:34:45.70&30:52:26.9&\nodata&\nodata&\nodata&\nodata&\nodata&Unknown&CS C19\\
432&01:34:45.91&30:53:04.4&19.70&\nodata&0.55&6.86&2.31&Cluster&CBF 84; Ma 2002b\\
433&01:34:45.99&30:50:50.4&\nodata&\nodata&\nodata&\nodata&\nodata&Stellar&CS U31\\
434&01:34:46.51&30:44:30.8&\nodata&\nodata&\nodata&\nodata&\nodata&Stellar&CS C29\\
435&01:34:46.68&30:49:25.0&\nodata&\nodata&\nodata&\nodata&\nodata&Stellar&CS U34\\
436&01:34:46.80&30:49:16.1&18.94&0.40&0.62&\nodata&\nodata&Cluster&CS U26\\
437&01:34:47.53&30:47:29.0&18.94&0.44&\nodata&\nodata&\nodata&Stellar&CS U43\\
438&01:34:49.62&30:21:55.5&16.11&0.81&\nodata&9.26&5.75&Cluster&CS C39; MD 52; Ma 2002a; Ma 2004b\\
439&01:34:50.10&30:47:04.1&16.55&0.26&0.57&8.06&4.69&Cluster&Hilt A; MD 56; CS M6; Ma 2002a; Ma 2004b\\
440&01:34:51.16&30:54:48.7&\nodata&\nodata&\nodata&\nodata&\nodata&Chip Gap&CS C14\\
441&01:34:52.24&30:50:05.6&\nodata&\nodata&\nodata&\nodata&\nodata&Cluster&CS U32\\
442&01:34:52.90&30:10:51.2&\nodata&\nodata&\nodata&\nodata&\nodata&Galaxy&MD 51\\
443&01:34:53.17&30:51:47.8&\nodata&\nodata&\nodata&\nodata&\nodata&Cluster&CS U15; CS U17\\
444&01:34:56.88&30:52:35.0&\nodata&\nodata&\nodata&\nodata&\nodata&Unknown&CS U16\\
445&01:34:58.32&30:31:10.5&\nodata&\nodata&\nodata&\nodata&\nodata&Unknown&CS U122\\
446&01:35:01.56&30:51:27.0&\nodata&\nodata&\nodata&\nodata&\nodata&Cluster&CS U18\\
447&01:35:04.69&30:46:10.6&\nodata&\nodata&\nodata&\nodata&\nodata&Cluster&CS U58\\
448&01:35:14.01&30:52:32.9&18.01&0.60&\nodata&\nodata&\nodata&Galaxy&CS U90\\
449&01:35:18.25&30:49:53.9&\nodata&\nodata&\nodata&\nodata&\nodata&Cluster&CS C22\\
450&01:35:45.70&30:26:51.4&17.20&0.82&\nodata&10.02&5.91&Galaxy&MD 57; Ma 2002a; Ma 2004b\\
451&01:36:05.40&30:58:19.7&\nodata&\nodata&\nodata&\nodata&\nodata&Galaxy&MD 58\\
\enddata
\tablecomments{Units of right ascension are hours, minutes, and seconds, and
units of declination are degrees, arcminutes, and arcseconds.}
\tablenotetext{a}{Units of age are in years}
\tablenotetext{b}{Units of mass are in solar masses}
\end{deluxetable}


\clearpage

\clearpage
\begin{figure}
\epsscale{0.7}
\plotone{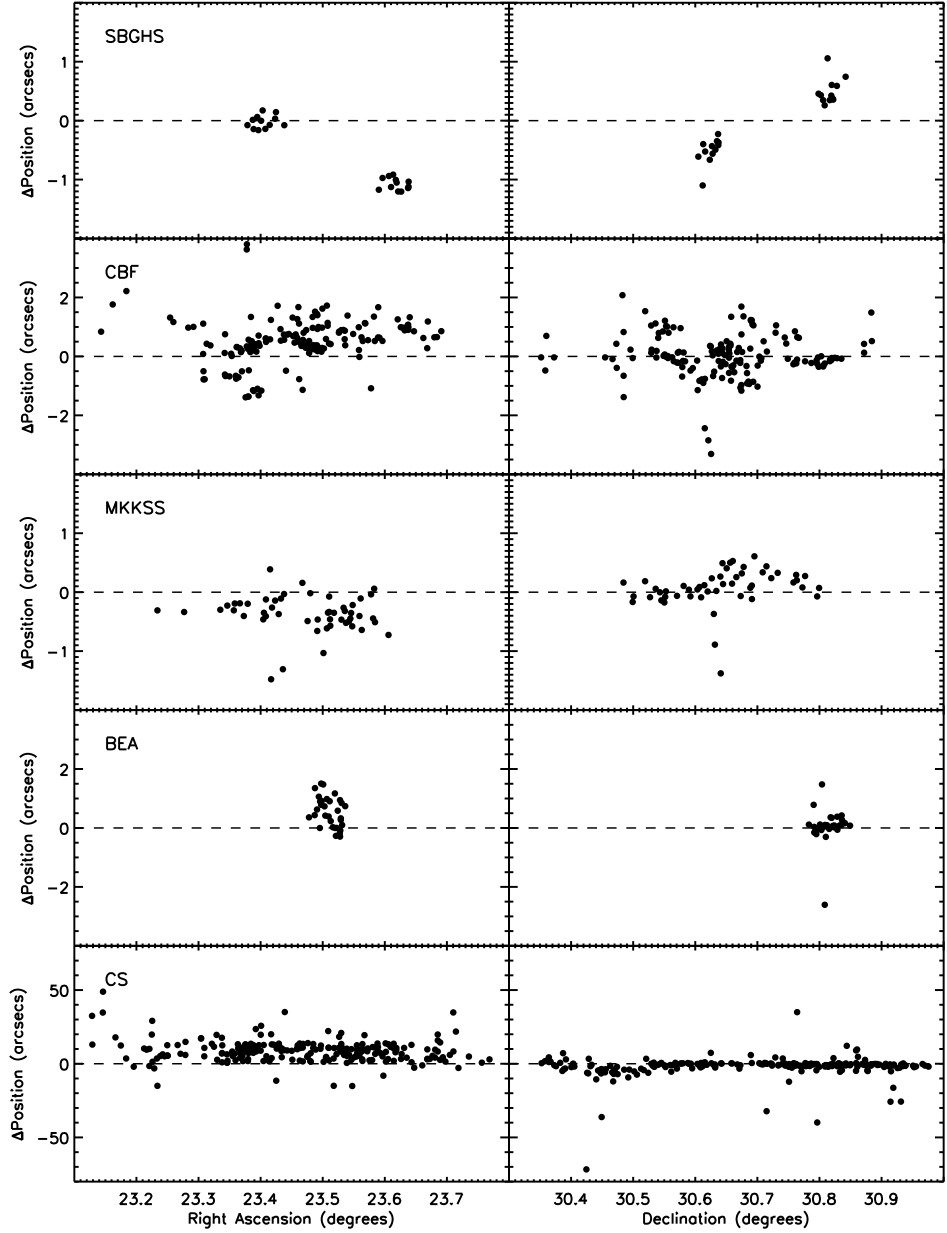}
\caption{Right ascension (left) and declination (right) differences between each
catalog and those measured from the Local Group Survey (Massey et al. 2006) 
images used in the present study.}
\end{figure}

\clearpage
\begin{figure}
\epsscale{0.8}
\plotone{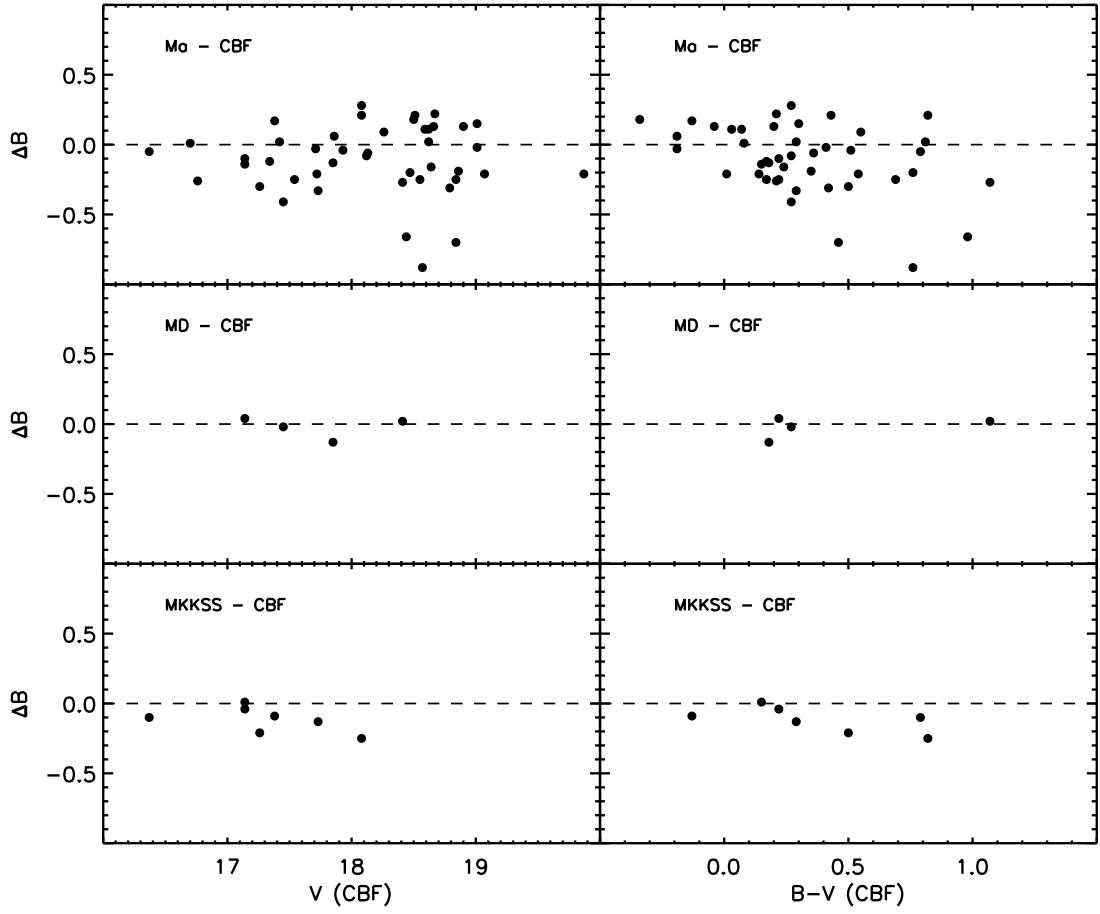}
\caption{The difference in B magnitude versus B (left) and B--V (right) for all papers with B and V photometry in common with CBF.}
\end{figure}
\clearpage
\begin{figure}
\epsscale{0.8}
\plotone{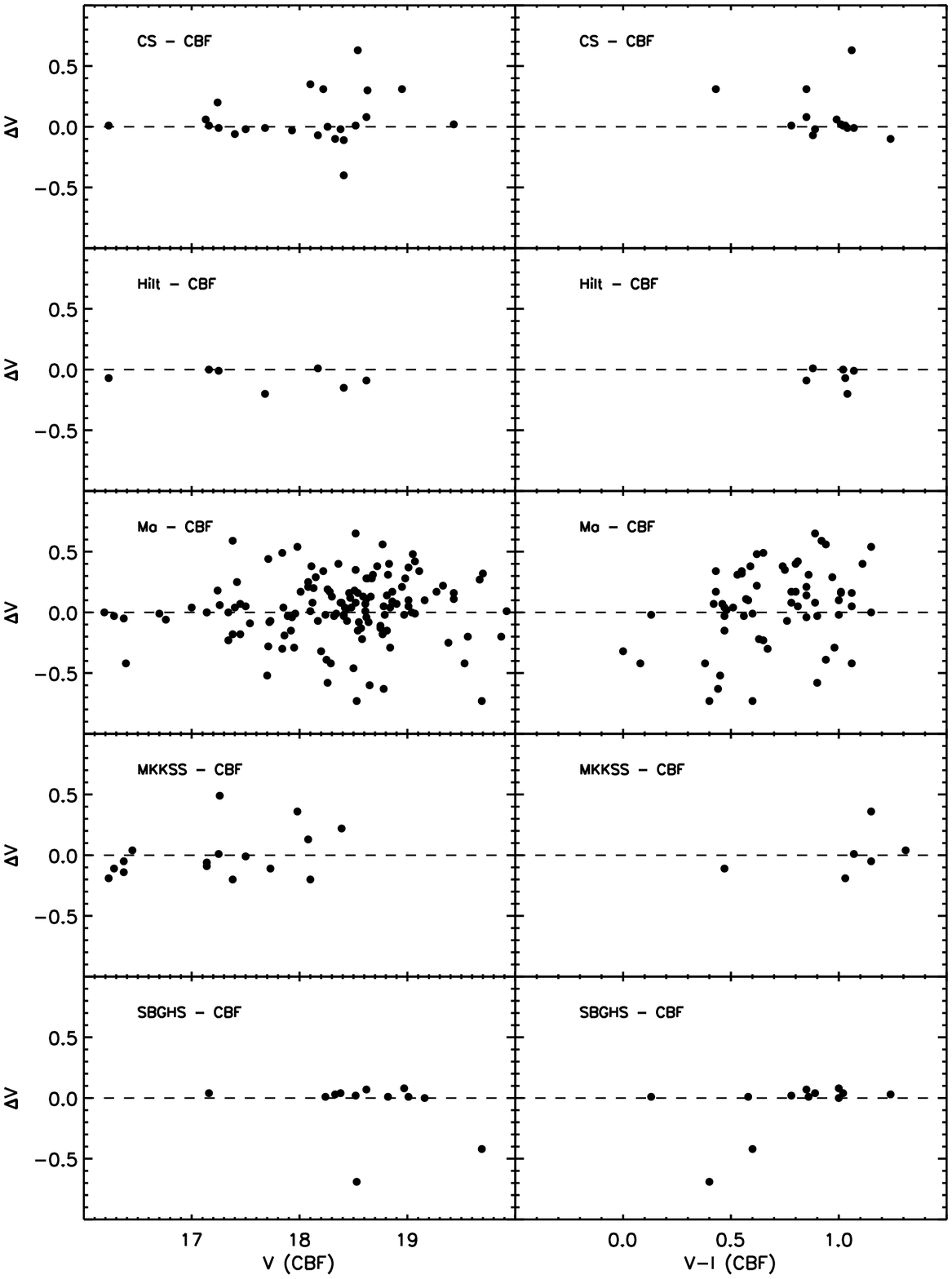}
\caption{The difference in V magnitude versus V (left) and V--I (right) for all papers with V and I photometry in common with CBF.}
\end{figure}
\clearpage
\begin{figure}
\epsscale{0.8}
\plotone{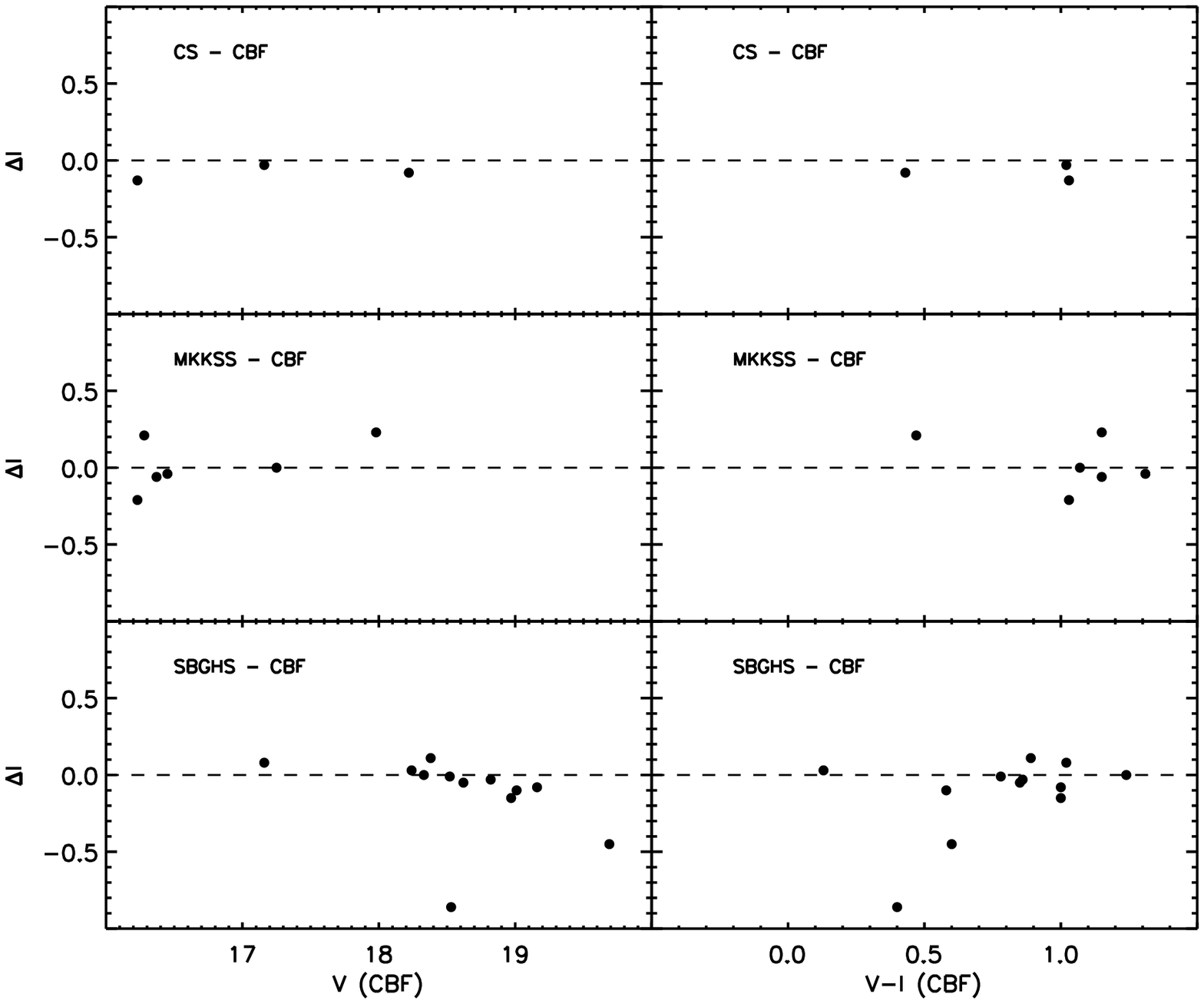}
\caption{The difference in I magnitude versus V (left) and V--I (right) for all papers with V and I photometry in common with CBF.}
\end{figure}

\clearpage
\begin{figure}
\epsscale{0.8}
\plotone{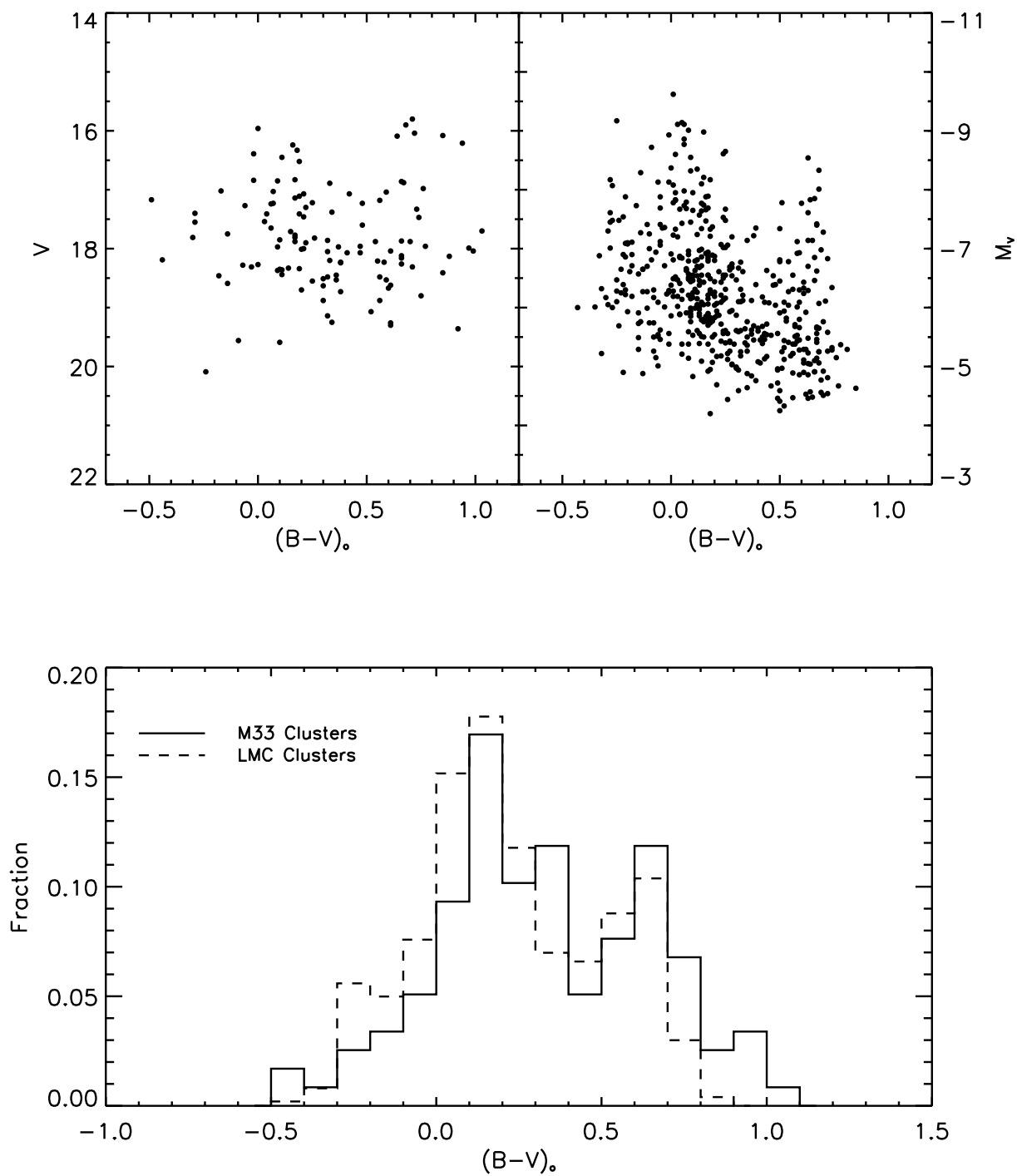}
\caption{The top panels show the cluster color-magnitude diagrams for M33 (left) using our
catalog and the Large Magellanic Cloud (right) from the catalog of Bica et al. (1999). 
A constant reddening correction of E(B--V)=0.1 has been applied to all clusters. The lower panel 
displays the color histograms of these populations scaled to unit area.}
\end{figure}

\clearpage
\begin{figure}
\epsscale{0.75}
\plotone{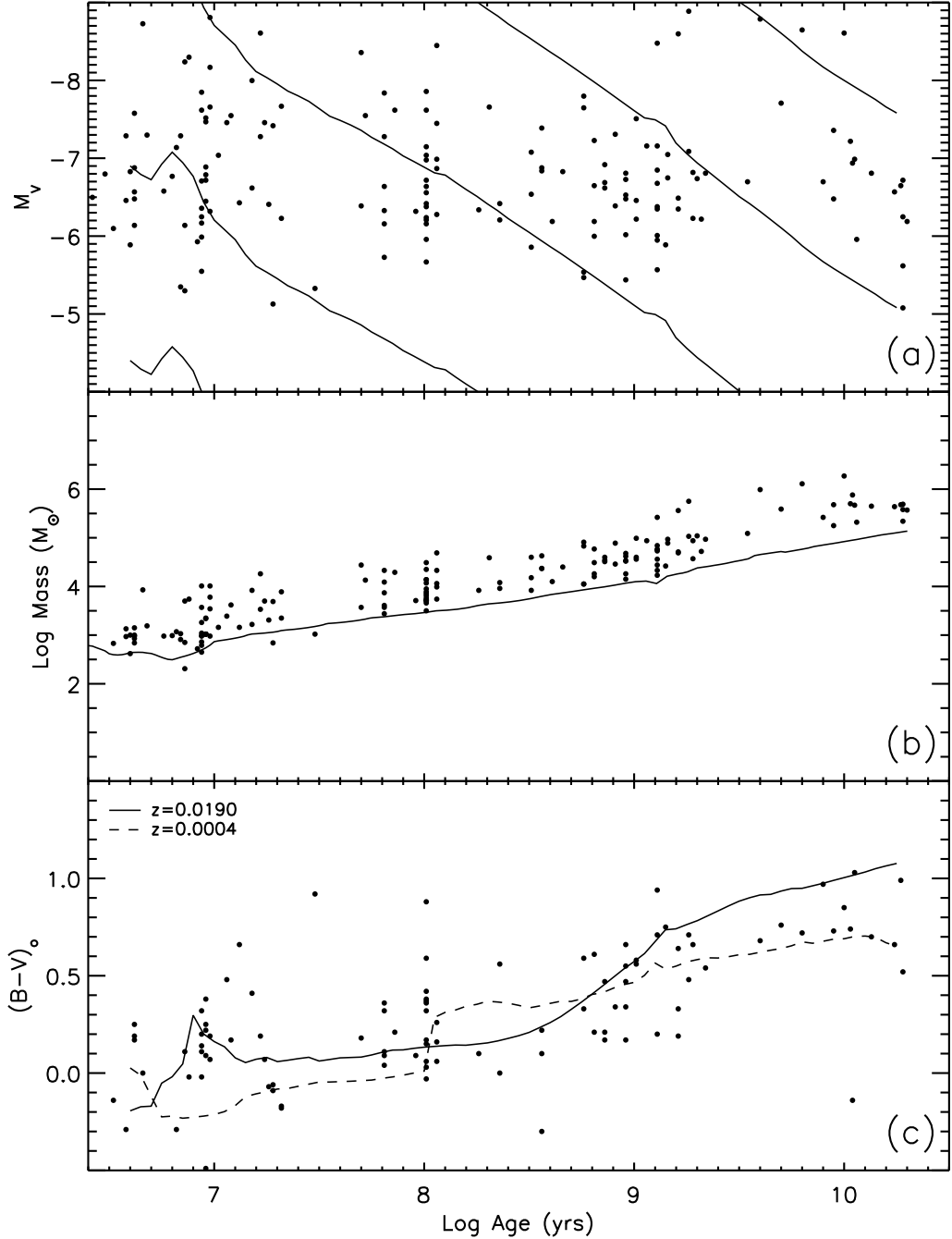}
\caption{(a) The absolute V magnitude for high confidence clusters as a function of their ages
as tabulated in our catalog.  Overplotted are theoretical lines corresponding to masses of
10\textsuperscript{2}, 10\textsuperscript{3}, 10\textsuperscript{4}, 
10\textsuperscript{5}, and 10\textsuperscript{6}$M_{\odot}$ from Girardi et al. (2002) assuming
a $M/L$ ratio of unity.  (b) Age versus Mass for high 
confidence clusters.  The masses are interpolated from the diagram above. The solid line
represents the fading line predicted by the Bruzual \& Charlot (2003) models for Z=0.008
shifted to match the lower envelope of points.
  (c) Dereddened color as a function of age for high 
confidence clusters in our catalog.  Overplotted are theoretical models for single
stellar populations from Girardi et al. (2002) for Z=0.0004 and Z=0.019.}
\end{figure}

\clearpage
\begin{figure}
\epsscale{0.9}
\plotone{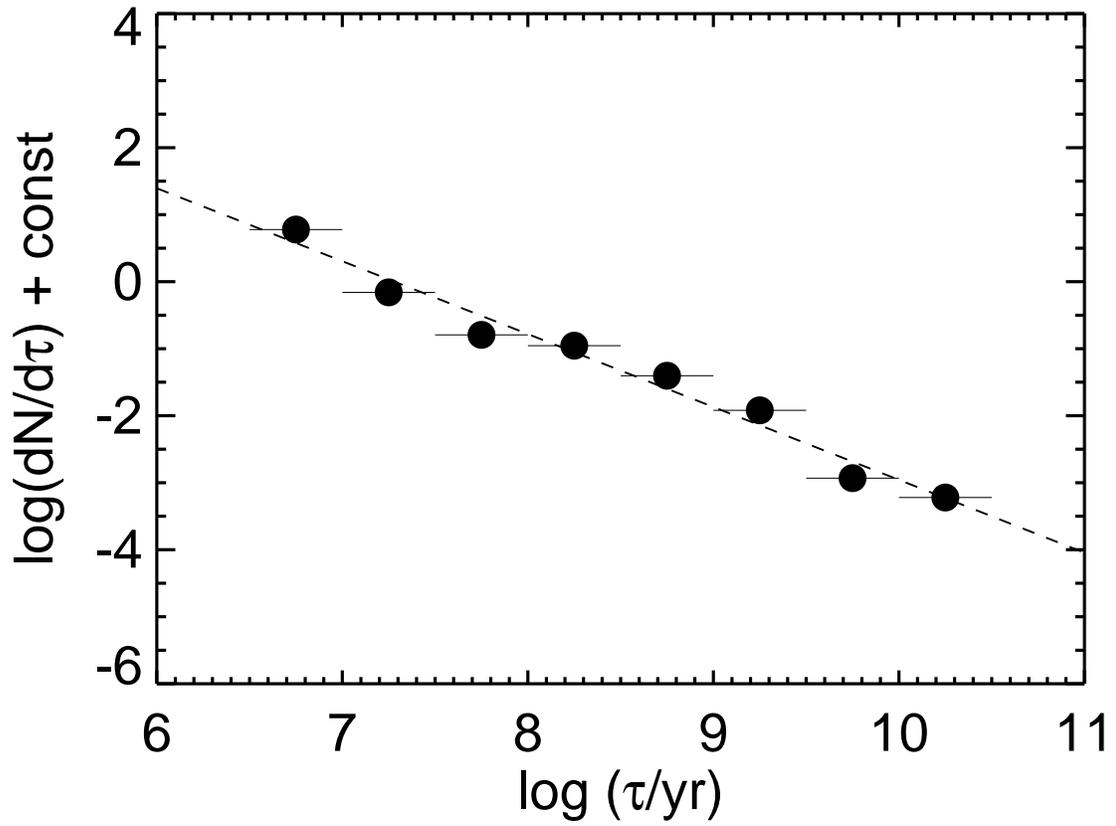}
\caption{The age distribution of star clusters in M33. This figure has been constructed
following the precepts of Chandar et al. (2007). The dashed line represents a
power law of the form $dN_{cluster}/d$$\tau$~$\propto$~$\tau$$^{\alpha}$
with a slope of --1.09$\pm$0.07.}
\end{figure}

\clearpage
\begin{figure}
\epsscale{0.8}
\plotone{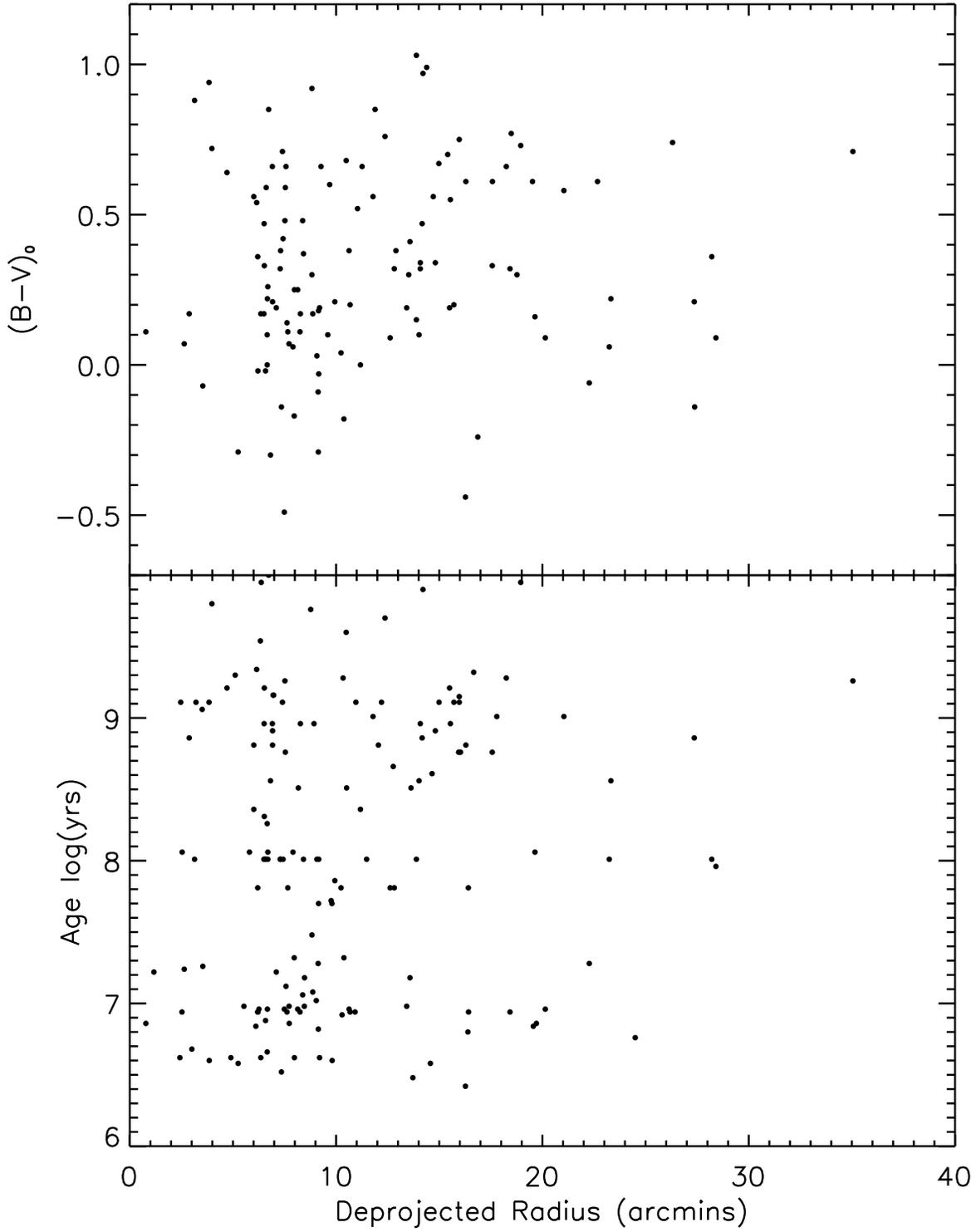}
\caption{The dereddened color (top) and cluster age (bottom) are plotted versus deprojected galactocentric radius for the 255 high confidence clusters.}
\end{figure}

\clearpage
\begin{figure}
\epsscale{0.8}
\plotone{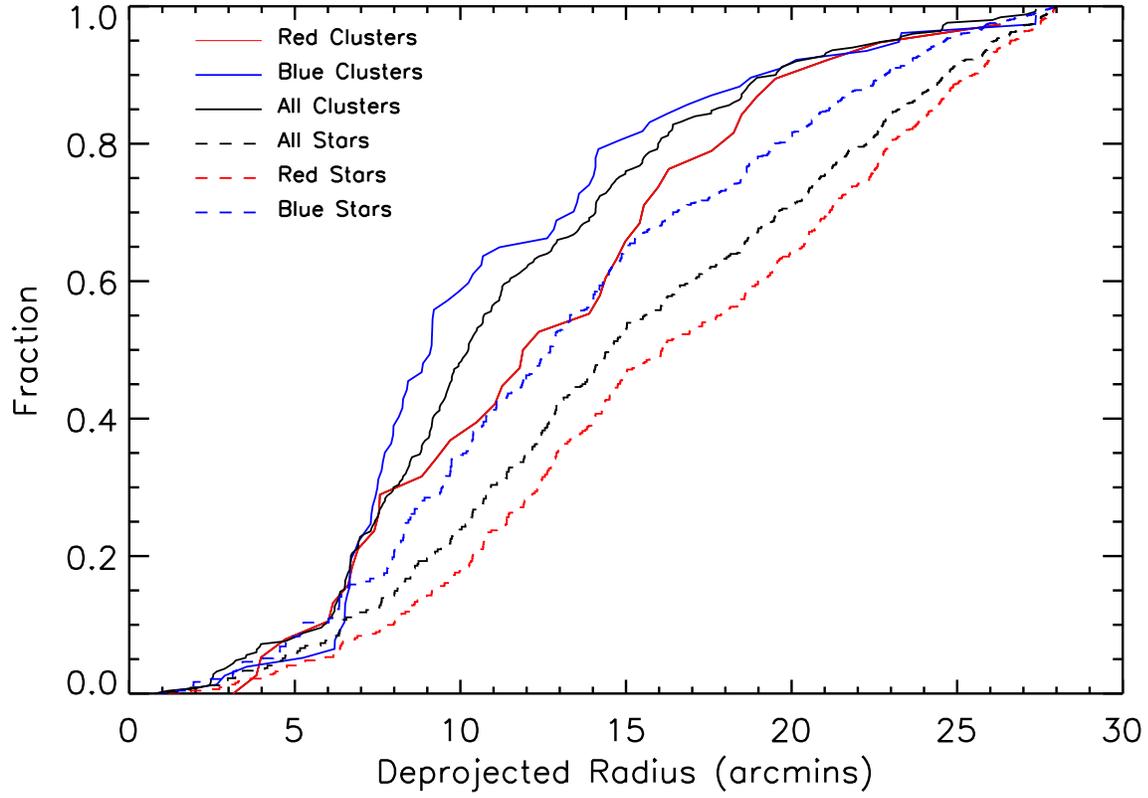}
\caption{Cumulative radial distributions for the star clusters and field stars in M33. 
A Kolmogorov-Smirnov (K-S) statistical analysis has been applied to these profiles.
See text for details.}
\end{figure}

\clearpage
\begin{figure}
\epsscale{0.8}
\plotone{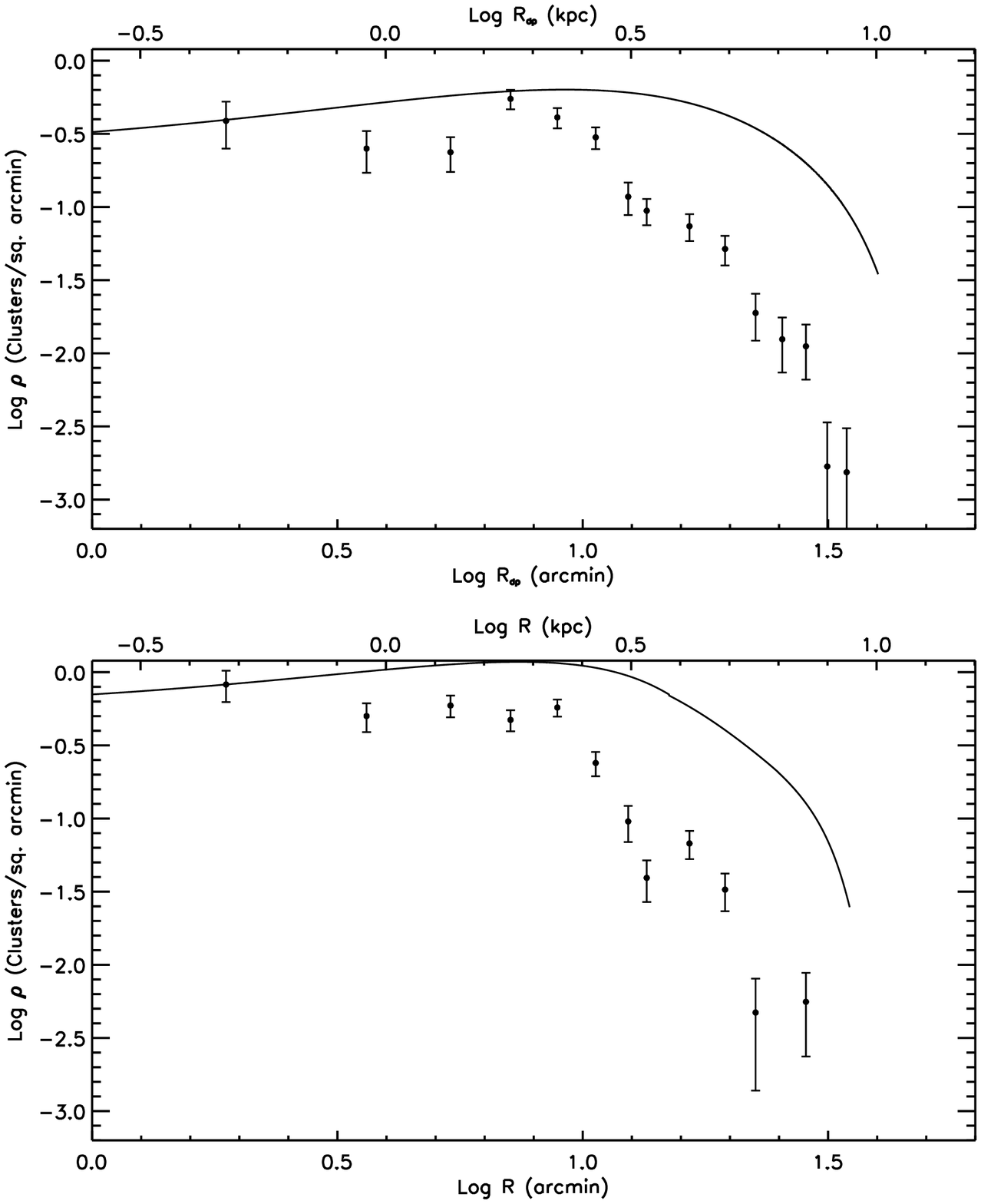}
\caption{Radial density profile of high confidence M33 clusters (filled circles) and field stars 
(solid line) from the
``M33 CFHT Variability Survey" of Hartman et al. (2006).  The upper panel shows the
deprojected radius while the lower panel displays the projected radius.}
\end{figure}

\clearpage




\begin{thebibliography}{}
\bibitem[]{595} Bedin, L. R., Piotto, G., Baume, G., Momany, Y., Carraro, G., Anderson, J.,
Messineo, M., \& Ortolani, S. 2005, \aap, 444, 831 (BEA)
\bibitem[]{597} Bica, E. L. D., Schmitt, H. R., Dutra, C. M., \& Oliveira, H. L. 1999, \aj, 117, 238
\bibitem[]{598} Bruzual, G., \& Charlot, S. 2003, \mnras, 344, 1000
\bibitem[]{599} Carraro, G., \& Chiosi, C. 1994, \aap, 288, 751
\bibitem[]{600} Chandar, R., Bianchi, L., \& Ford, H. C. 1999, \apjs, 122, 431 (CBF99)
\bibitem[]{601} Chandar, R., Bianchi, L., \& Ford, H. C. 2001, \aap, 366, 498 (CBF01)
\bibitem[]{602} Chandar, R., Bianchi, L., Ford, H. C., \& Sarajedini, A. 2002, \apj, 564,
712
\bibitem[]{604} Chandar, R., Fall, S. M., \& Whitmore, B. 2006, \apj, 650, L111
\bibitem[]{605} Christian, C. A. \& Schommer, R. C. 1982, \apjs, 49, 405 (CS)
\bibitem[]{606} Christian, C. A. \& Schommer, R. C. 1983, \apj, 275, 92 
\bibitem[]{607} Christian, C. A. \& Schommer, R. C. 1988, \aj, 95, 704
\bibitem[]{608} Fall, S. M., Chandar, R., \& Whitmore, B. C. 2005, \apj, 631, L133
\bibitem[]{609} Fan, X. et al. 1996, \aj, 112, 628
\bibitem[]{610} Galleti, S., Bellazzini, M., \& Ferraro, F. R. 2004, \aap, 423, 925
\bibitem[]{611} Girardi, L., Bressan, A., Bertelli, G., \& Chiosi, C. 2000, \aaps, 141, 371
\bibitem[]{612} Girardi, L., Bertelli, G., Bressan, A., Chiosi, C., Groenewegen, M. A. T.,
Marigo, P., Salasnich, B., \& Weiss, A. 2002, \aap, 391, 195
\bibitem[]{614} Grocholski, A. J., Sarajedini, A., Olsen, K. A. G., \&
Tiede, G. P. 2007, \aj, submitted
\bibitem[]{616} Hartman, J. D., Bersier, D., Stanek, K. Z., Beaulieu, J. -P., 
Kaluzny, J., Marquette, J. -B., Stetson, P. B., \& Schwarzenberg-Czerny, A. 2006,
\mnras, 371, 1405
\bibitem[]{619} Hiltner, W. A. 1960, \apj, 131, 161 (Hilt)
\bibitem[]{620} Hunter, D. A., Elmegreen, B. G., Dupuy, T. J., \& Mortonsn, M., 2003,
\aj, 126, 1836
\bibitem[]{622} Kim, M., Kim, E., Lee, M. G., Sarajedini, A., \& Geisler, D. 2002, \aj,
123, 244
\bibitem[]{} Landolt, A. U. 1983, AJ, 88, 439
\bibitem[]{} Landolt, A. U. 1992, AJ, 104, 340
\bibitem[]{624} Ma, J., Zhuo, X., Wu, H., Chen, J., Jiang, Z., Zhu, J., \& Xue, S.
2001, \aj, 122, 1796 (Ma01)
\bibitem[]{626} Ma, J., Zhou, X., Chen, J. -S., Wu, H., Jiang, Z. -J., Xue, S. -J.,
\& Zhu, J. 2002a, ChJAAp, 2, 197 (Ma02a)
\bibitem[]{628} Ma, J., Zhou, X., Chen, J., Wu, H., Jiang, Z., Xue, S., \& Zhu, J.
2002b, \aj, 123, 3141 (Ma02b)
\bibitem[]{630} Ma, J., Zhou, X., Chen, J., Wu, H., Kong, X., Jiang, Z., Zhu, J., \&
Xue, S. 2002, AcA, 52, 453 (Ma02c)
\bibitem[]{632} Ma, J., Zhou, X., Chen, J. 2004, \aap, 413, 563 (Ma04a)
\bibitem[]{633} Ma, J., Zhou, X., \& Chen, J. -S. 2004, ChJAAp, 4, 125 (Ma04b)
\bibitem[]{634} Massey, P., Olsen, K. A. G., Hodge, P. W., Strong, S. B., 
Jacoby, G. H., Schlingman, W., \& Smith, R. C. 2006, AJ, 131, 2478
\bibitem[]{636} Melnick, J., \& D'odorico, S. 1978, \aaps, 34, 249 (MD)
\bibitem[]{637} Mochejska, B. J., Kaluzny, J., Krockenberger, M., Sasselov, D. D.,
\& Stanek, K. 1998, AcA, 48, 455 (MKKSS)
\bibitem[]{639} Rafelski, M., \& Zaritsky, D. 2005, \aj, 129, 2701
\bibitem[]{640} Regan, M. W. \& Vogel, S. N. 1994, \apj,  434, 536
\bibitem[]{641} Reiss, A. 2003, ACS-ISR 2003-09
\bibitem[]{646} Sarajedini, A., Barker, M. K., Geisler, D., Harding, P., \&
Schommer, R. 2007, AJ, 133, 290 (SBGHS)
\bibitem[]{648} Schommer, R. A., Christian, C. A., Caldwell, N., Bothun, G. D.,
\& Huchra, J. 1991, \aj, 101, 873
\bibitem[]{650} Tasker, E. J., \& Bryan, G. L. 2006, \apj, 641, 878
\bibitem[]{651} Tasker, E. J., \& Bryan, G. L. 2007, in preparation
\bibitem[]{652} Whitmore, B. C., Chandar, R. \& Fall, S. M. 2007, \aj, 133, 1067
\bibitem[]{653} Wielen, R. 1977, \aap, 60, 263
\bibitem[]{654} Wielen, R., Fuchs, B., \& Dettbarn, C. 1996, \aap, 314, 438
\end{thebibliography}
\end{document}